\RequirePackage{fix-cm}
\documentclass[twocolumn,epjc3]{ARXIVsvjour3}  
\smartqed  
\RequirePackage{graphicx}

\usepackage{amsmath,amssymb,amsfonts}
\usepackage{graphicx}
\usepackage{xcolor}
\usepackage{mathrsfs}
\usepackage{hyperref}
\usepackage{float}
\usepackage{lipsum}
\usepackage{mathtools}
\usepackage{cuted}
\hypersetup{colorlinks,citecolor=blue,urlcolor=blue,linkcolor=blue}
\newcommand\beq{\begin{eqnarray}}
\newcommand\eeq{\end{eqnarray}}

\journalname{Eur. Phys. J. C}

\begin{document}

\title{Criteria for projected discovery and exclusion sensitivities of counting
experiments}

\author{Prudhvi N. Bhattiprolu\thanksref{addr1}
        \and
        Stephen P. Martin\thanksref{addr1}
        \and
        James D. Wells\thanksref{addr2}
}

\institute{Northern Illinois University, DeKalb IL 60115, USA\label{addr1}
           \and
           University of Michigan, Ann Arbor, MI 48109, USA\label{addr2}
           }

\date{Received: date / Accepted: date}

\maketitle

\begin{abstract}
The projected discovery and exclusion capabilities of particle physics and astrophysics/cosmology experiments are often quantified using the median expected $p$-value or its corresponding significance. We argue that this criterion leads to flawed results, which for example can counterintuitively project lessened sensitivities if the experiment takes more data or reduces its background. We discuss the merits of several alternatives to the median expected significance, both when the background is known and when it is subject to some uncertainty. We advocate for standard use of the ``exact Asimov significance" $Z^{\rm A}$ detailed in this paper.
\end{abstract}

\section{Introduction}
\label{intro}

Consider the problem of assessing the efficacy of a planned experiment that will measure event counts that could be ascribed either to a new physics signal or a standard physics background. The criteria for discovery or exclusion of the signal can be quantified in terms of the $p$-value. In general, for a given experimental result, $p$ is the probability of obtaining a result of equal or greater incompatibility with a null hypothesis $H_0$. In high-energy physics searches, for example, the one-sided $p$-value results are usually reported in terms of the significance
\beq
Z &=& \sqrt{2}\, {\rm erfc}^{-1}(2p),
\label{eq:Zfromp}
\eeq
and the criteria for discovery and exclusion have often been taken, somewhat arbitrarily, as $Z>5$ ($p< 2.867\times 10^{-7}$) and $p<0.05$ $(Z>1.645)$, respectively.

Here, we suppose for simplicity that both signal and background are governed by independent Poisson statistics with means $s$ and $b$ respectively, where $s$ is known and $b$ may be subject to some uncertainty. For assessing the prospects for discovery, one simulates many equivalent 
pseudo-experiments with data generated
under the assumption $H_{\rm data} = H_{s+b}$ that both signal and background are present, obtaining observed events $n_1,n_2,n_3,\ldots$. One then calculates the $p$-value for each of those simulated experiments ($p_1,p_2,p_3,\ldots$) with respect to the null hypothesis $H_0 = H_{b}$ that only background is present. For exclusion, the roles of the two hypotheses are reversed; the pseudo-experiment
data is generated under the assumption $H_{\rm data} = H_b$ that only background is present,
and the null hypothesis $H_{0} = H_{s+b}$ is that both signal and background are present, so that 
a different set of $p$-values is obtained. 
The challenge is to synthesize the results 
in the limit of a very large number of pseudo-experiments into a significance estimate $Z_{\rm disc}$
or $Z_{\rm excl}$. There is no agreement on this step, which is the primary focus of this paper. 

A common measure~\cite{CowanPDG:2018} of the power of an experiment is the median expected significance $Z^{\rm med}$ for discovery or exclusion of some important signal (i.e., the median of $Z(p_1),Z(p_2),Z(p_3),\ldots$ for the simulated $p$-values). A reason to use the median (rather than mean) is that eq.~(\ref{eq:Zfromp}) is non-linear, so that the mean of a set of $Z$-values is not the same as the $Z$-value of the corresponding mean of $p$-values. 

However, $Z^{\rm med}$ has a counter-intuitive flaw, which is most prominent when $s$ and $b$ are not too large, and especially for exclusion. As we show 
in the following examples, for a given fixed $s$, $Z^{\rm med}$ can actually significantly {\em increase} as $b$ increases. Similarly, for a given fixed $b$, $Z^{\rm med}$ can decrease as $s$ is increased. This leads to the paradoxical situation that an experiment could be judged worse, according to the $Z^{\rm med}$ criteria, if it acquires more data, or if it reduces its background. In this paper, we discuss this problem, and consider some alternatives to $Z^{\rm med}$.

\section{Known background case}

The Poisson probability of observing $n$ events, given a mean $\mu$, is
\beq
P(n|\mu) &=& e^{-\mu} \mu^n/n!
.
\label{eq:Poissonprobability}
\eeq 
Consider first the idealized case that the signal and background Poisson means $s$ and $b$ are both known exactly. One can then generate pseudo-experiment results for $n$, using $\mu = s+b$ for the discovery case, and $\mu = b$ for the exclusion case. A large number of simulated pseudo-experiments can be generated randomly via Monte Carlo simulation methods, as described in the Introduction. However, for all cases in this paper, it is equivalent but much more efficient and accurate to consider exactly once each result $n$ that can contribute non-negligibly, and then weight the results according to the probability of occurrence.

The $p$-value for discovery, if $n$ events are observed, is
\beq
p_{\rm disc}(n,b) = \sum_{k=n}^\infty P(k|b) = 
\gamma(n,b)/\Gamma(n)
,
\label{eq:pdiscnb}
\eeq
while that for exclusion is
\beq
p_{\rm excl}(n,b,s) = \sum_{k=0}^n P(k|s+b) = 
\frac{\Gamma(n+1, s+b)}{\Gamma(n+1)} ,
\label{eq:pexclnbs}
\eeq
where $\Gamma(x)$, $\gamma(x,y)$, and $\Gamma(x,y)$ are the ordinary, lower incomplete, and upper incomplete gamma functions, respectively. The
median $p$-value among the pseudo-experiments can now be converted, using eq.~(\ref{eq:Zfromp}),
to obtain $Z^{\rm med}_{\rm disc}(s,b)$ and $Z^{\rm med}_{\rm excl}(s,b)$. 

\begin{figure*}
 \includegraphics[width=8.1cm]{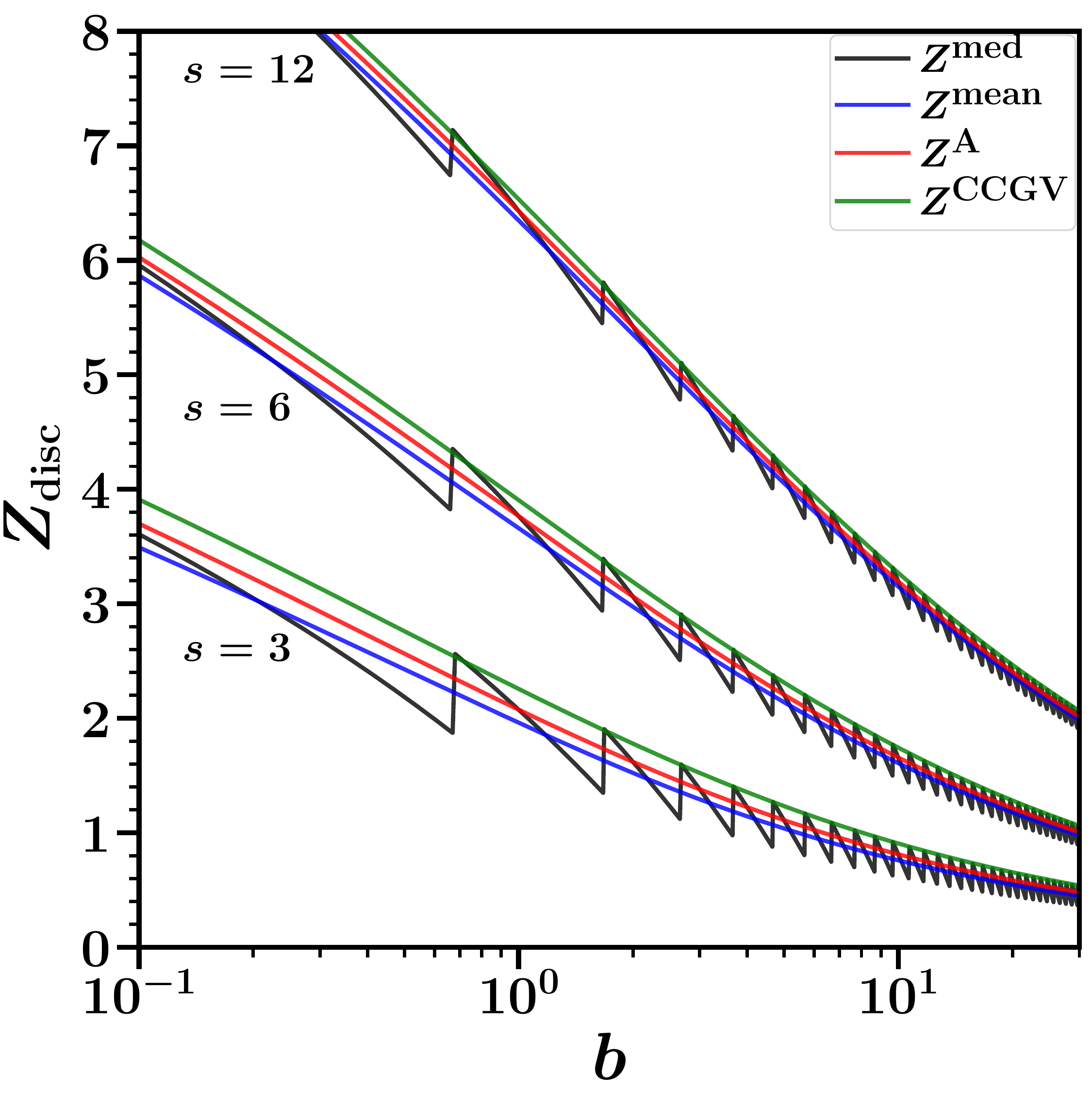}%
  \hspace{0.7cm}
 \includegraphics[width=8.35cm]{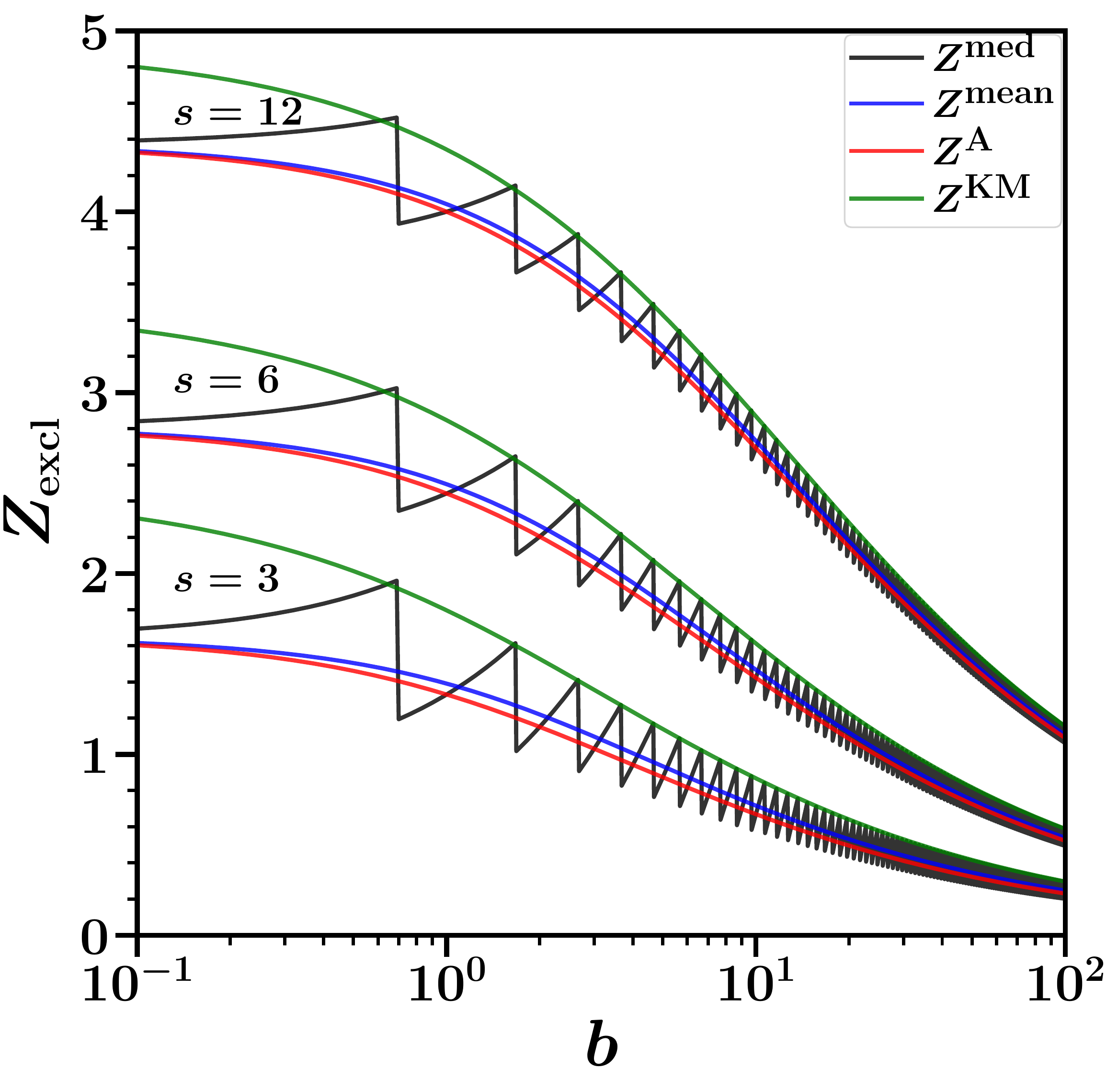}%
 \caption{Expected significances for discovery (left) and exclusion (right), for signal means $s = 3$, 6, and 12, as functions of the background mean $b$. Shown are $Z^{\rm med}$, $Z^{\rm mean}$, $Z^{\rm A}$, and the approximations $Z^{\rm CCGV}$ and $Z^{\rm KM}$ from refs.~\cite{Cowan:2010js,Cowan} 
and \cite{Kumar:2015tna}.
The median expected significances show a sawtooth behavior, rather than decreasing monotonically with $b$.
\label{fig:deltabeq0}}
\end{figure*}

Some typical results for $Z_{\rm disc}^{\rm med}$ and
$Z_{\rm excl}^{\rm med}$ as a function of $b$ are shown in Figure \ref{fig:deltabeq0}.
They each have a ``sawtooth" shape, rather than monotonic as one might perhaps expect. This illustrates the unfortunate feature mentioned in the Introduction that the median expected $Z$ can increase with increasing $b$. As noted in \cite{Cowan:2010js,Cowan} for $Z_{\rm disc}^{\rm med}$, the underlying reason is that the allowed values of $n$ are discrete (integers), causing the median to remain at a fixed value instead of varying continuously in response to changes in $s$ or $b$. We emphasize that this sawtooth behavior is exactly reproducible for any sufficiently large number of pseudo-experiments, and has nothing to do with randomness from insufficient sampling. It is more prominent for exclusion than for discovery, because the number of events relevant for the median pseudo-experiment is smaller. Also, note that for larger $b$, the sawteeth get closer together as the integer $n$ of the median gets larger, but the height of the sawtooth envelope remains significant. This is effectively a sort of practical randomness in $Z^{\rm med}$, as tiny changes in $s$ or $b$ will move one between the top and the bottom of the sawtooth envelope.

We now consider several alternatives to $Z^{\rm med}$. First, one can take the  arithmetic mean of the $Z$-values directly, which we call $Z^{\rm mean}$. 
(In computing $Z^{\rm mean}_{\rm disc}$, we use $Z=0$ for no observed events, $n=0$. A reasonable alternative definition for both $Z^{\rm mean}_{\rm disc}$ and $Z^{\rm mean}_{\rm excl}$ would be to use $Z=0$ for all outcomes $n$ that give a negative $Z$. That would give slightly larger values for $Z^{\rm mean}$, but usually negligibly so except when
$Z^{\rm mean}$ is uninterestingly small anyway.) 
Second, one can take the arithmetic mean of the $p$-values, and then convert these to $Z$ values, which we call $Z^{p\>\rm mean}$. Third, one can consider the $Z$-value obtained for the mean $n$ (i.e., average over the simulated $n_1,n_2,n_3,\ldots$); 
the use of the mean data for computing the expected significance has been used in 
\cite{Bartsch:2005xxa,Aad:2009wy} and \cite{Cowan:2010js,Cowan}
and was called the Asimov data in the latter three references. Refs.~\cite{Cowan:2010js,Cowan} obtained
an Asimov approximation to $Z^{\rm med}_{\rm disc}$:
\beq
Z_{\rm disc}^{\rm CCGV} = \sqrt{2[(s+b)\ln(1+s/b) -s]},
\label{eq:ZAsimovCowan}
\eeq
and ref.~\cite{Kumar:2015tna} gave a similar result for exclusion:
\beq
Z_{\rm excl}^{\rm KM} = \sqrt{2[s - b\ln(1+s/b)]}
.
\label{eq:ZAsimovKM}
\eeq
These are both based on a likelihood ratio method approximation
(valid in the limit of a large event sample) for $Z$ given in \cite{LiMa:1983} 
in the context of $\gamma$-ray astronomy.
They both approach the familiar but cruder approximation $s/\sqrt{b}$, but only in the limit of very large $b$.

In this paper, we propose instead to simply use for the Asimov approximation the exact $p$-values in eqs.~(\ref{eq:pdiscnb}) and (\ref{eq:pexclnbs}) with $n$ replaced by its expected means: 
\beq
\langle n_{\rm disc} \rangle = s+b,
\qquad
\langle n_{\rm excl} \rangle = b, 
\label{eq:nmeanDeltabeqzero}
\eeq
so that 
\beq
p^{\rm Asimov}_{\rm disc} &=& \gamma(s+b,b)/\Gamma(s+b),
\label{eq:pAsimovdiscDeltabeqzero}
\\
p^{\rm Asimov}_{\rm excl} &=& \Gamma(b+1,s+b)/\Gamma(b+1),
\label{eq:pAsimovexclDeltabeqzero}
\eeq
which can be readily converted to $Z$-values using eq.~(\ref{eq:Zfromp}). We call this the ``exact Asimov significance" and denote it by $Z^{\rm A}$.

Along with $Z^{\rm med}$, Figure \ref{fig:deltabeq0} also shows $Z^{\rm mean}$ and $Z^{\rm A}$ for the discovery
and exclusion cases, together with $Z^{\rm CCGV}_{\rm disc}$, and $Z^{\rm KM}_{\rm excl}$, as a function of $b$, for fixed $s=3,6,12$. 
Both $Z^{\rm mean}$ and $Z^A$ are within the $Z^{\rm med}$ sawtooth envelopes, but decrease monotonically with $b$. We conclude that they are both sensible measures
of the expected significance. In the discovery case, $Z^{\rm mean}$ is generally slightly more conservative than $Z^{\rm A}$, and the reverse is true for the exclusion case. The previously known Asimov approximations $Z^{\rm CCGV}_{\rm disc}$ and $Z^{\rm KM}_{\rm excl}$ of refs.~\cite{Cowan,Cowan:2010js} and \cite{Kumar:2015tna} are considerably less conservative, lying near the upper edges of the $Z^{\rm med}$ sawtooth envelopes. 

Not shown in 
Fig.~\ref{fig:deltabeq0}
is $Z^{p\, \rm mean}$, which we find is much lower than all of the others, due to being dominated by unlikely outcomes with large $p$-values, and therefore not a reasonable measure of the expected significance. 
Although we do not recommend its use, we note the amusing fact 
$Z^{p\, \rm mean}_{\rm disc} = Z^{p\, \rm mean}_{\rm excl}$, the proof of which does not rely on the assumed probability distribution, and so also holds exactly in the case of an uncertain background discussed below.

One sometimes sees $s/\sqrt{b}$ used as an estimate, but this is much larger than the $Z$'s shown in 
Fig.~\ref{fig:deltabeq0},
and, as is well-known, is not a good estimate of the significance for discovery or exclusion except when $b$ is large.

\begin{figure*}
 \includegraphics[width=8.0cm]{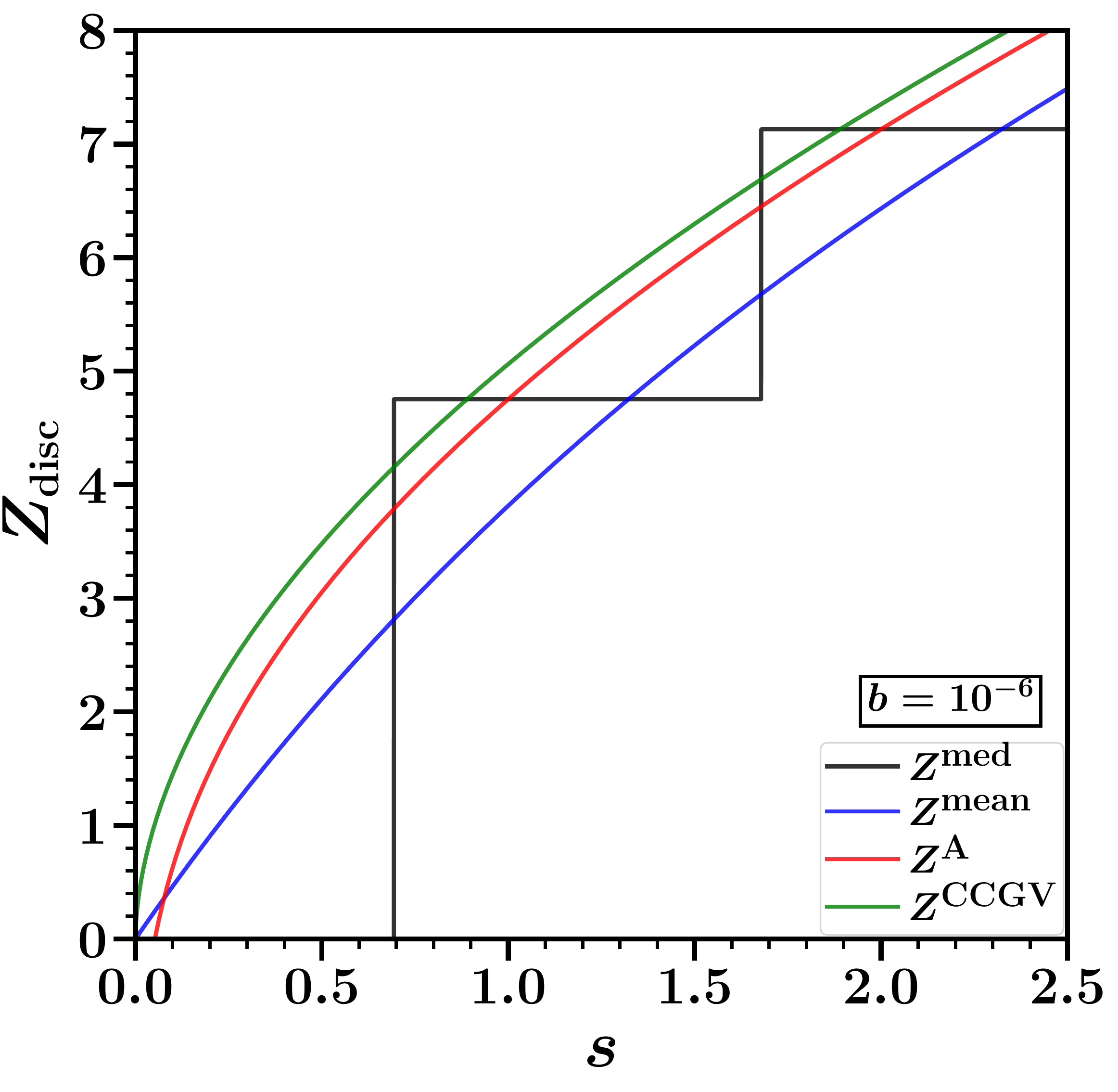}%
  \hspace{0.7cm}
 \includegraphics[width=8.0cm]{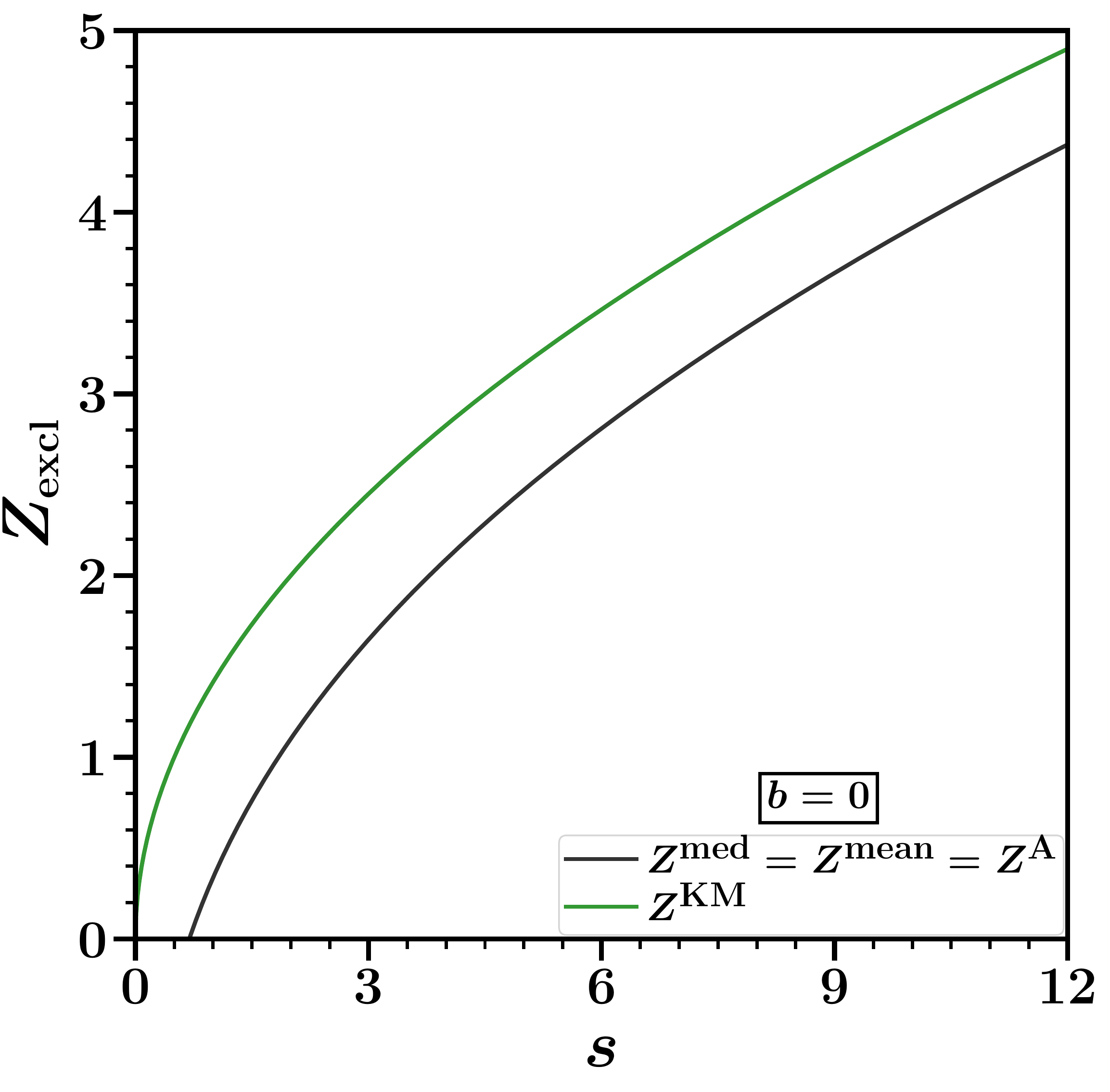}%
 \caption{Expected significances for discovery for an extremely small background mean
$b = 10^{-6}$ (left),
and exclusion for the strict limit $b=0$ (right), as functions of the signal mean $s$. Shown are $Z^{\rm med}$, $Z^{\rm mean}$, $Z^{\rm A}$, and the approximations $Z^{\rm CCGV}$ and $Z^{\rm KM}$ from refs.~\cite{Cowan:2010js,Cowan} 
and \cite{Kumar:2015tna}.
\label{fig:limitbeq0}}
\end{figure*}

We close this section by considering the extreme no-background limit $b \rightarrow 0$, with varying $s$. 
Background predictions much smaller than 1 can realistically come about from extrapolations
from other regions.
For discovery, if one takes $b=0$ exactly, then the significance for every pseudo-experiment is either zero 
(the value we have chosen to assign if no events are observed) 
or infinite (if even one event is observed). Since any non-zero $s$ would provide a 
non-zero mean number of events, one obtains $Z^A = \infty$ for all $s$ in that case. The limit $b \rightarrow 0$ in eq.~(\ref{eq:ZAsimovCowan}) is also seen to give $Z^{\rm CCGV} = \infty$. For the median expected significance, we instead get an infinitely large sawtooth. This is because the median number of events is 0 if 
$s < \ln(2)$, resulting in $Z^{\rm med} = 0$, and is at least 1 for
all $s > \ln(2)$, resulting in $Z^{\rm med} = \infty$. Therefore, for the discovery case it is perhaps more interesting to take $b$ extremely small, but non-zero. 
In Fig.~\ref{fig:limitbeq0}(a), 
we show $Z^{\rm med}$, $Z^{\rm mean}$, $Z^A$,  and $Z^{\rm CCGV}$, all for the
case $b= 10^{-6}$, as an example of a non-zero but very small expected background. The median number of events in the pseudo-experiments is $n=0$ for
$0 < s \leq s_1$, and is $n=1$ for $s_1 \leq s \leq s_2$, where 
$s_1 \approx \ln(2) \approx 0.693$
and $s_2 \approx 1.678$ is close to the solution of $s = \ln(2) + \ln(1+s)$.

In contrast,  in the exclusion case there is nothing singular for $b=0$. In particular, eq.~(\ref{eq:ZAsimovKM}) gives $Z^{\rm KM}_{\rm excl}(b=0) = \sqrt{2s}$. In each exclusion pseudo-experiment, the number of events observed is always $n=0$ because $b=0$, so that 
$Z^{\rm med}_{\rm excl} = Z^{\rm mean}_{\rm excl} = Z^{\rm A}_{\rm excl} = \sqrt{2} {\rm erfc}^{-1} (2 e^{-s})$ are all obtained from $p_{\rm excl} = \Gamma(1,s) = e^{-s}$.
These results are illustrated in Figure \ref{fig:limitbeq0}(b), which shows that
the estimate $Z^{\rm KM}_{\rm excl}$ in this extreme limit is larger than the others.
We note if $b=0$, then $s>2.996$ is needed to given an expected
95\% confidence level exclusion
$Z^{\rm mean}_{\rm excl} = Z^{\rm med}_{\rm excl} = 
Z^{\rm A}_{\rm excl} > 1.645$.

\section{Uncertain background case}

More realistically, the expected mean number of background counts can be subject to uncertainties of various sorts. In high-energy physics, the background uncertainty for a future experiment is often dominated by limitations in perturbative theoretical calculations or systematic effects, both of which are unknown (and indeed difficult to rigorously define) but can be roughly estimated or conjectured. There are also statistical uncertainties that will arise from a limited number of events in control or sideband regions. Here, we will consider, in part as a proxy for other types of uncertainties, the ``on-off problem" (see for example \cite{LiMa:1983,Gehrels:1986mj,ZhangRamsden:1990,Alexandreas:1993,Linnemann:2003vw,Cousins:2008zz}), in which the background is estimated by a measurement of $m$ Poisson events in a supposed background-only (off) region. The ratio of the background Poisson mean in this region to the background mean in the signal (on) region is assumed to be a known number $\tau$. 
It would also be interesting to consider the case of an uncertainty in $\tau$ itself, but that is beyond the scope of the present paper.
The point estimates for the Poisson mean and the uncertainty of the background in the signal region are then
\beq
\hat b = m/\tau,\qquad \Delta_{\hat b} = \sqrt{m}/\tau.
\label{eq:tradebhatDeltamtau}
\eeq
While this Poisson variance is certainly not a rigorous model for systematic or perturbative calculation uncertainties, we propose that it can also be used as a rough proxy for them, in the sense that a 
proposed estimate for $\hat b$ and $\Delta_{\hat b}$ can be traded for $(m,\tau)$ in the on-off problem.

We now assign probabilities $\Delta P$ to each possible count outcome $n$ in the on region, given $m$ events in the off region, following a hybrid Bayesian-frequentist approach 
by averaging \cite{Haines:1986ii,Alexandreas:1993,Cousins:1991qz,Linnemann:2003vw,Cousins:2008zz}
over the possible background means using a Bayesian posterior with a flat prior, 
\beq
P(b|m,\tau) = \tau (\tau b)^m e^{-\tau b}/m!,
\label{eq:posteriorPb}
\eeq
(normalized so that $\int_0^\infty \! db\> P(b|m,\tau) = 1$), from which we then find
\beq
\Delta P(n,m,\tau,s) = \int_0^\infty \!\!\! db\> P(b|m,\tau)\>  e^{-(s+b)} \frac{(s + b)^n}{n!}
\nonumber \phantom{xxx}&&
\\
= \frac{\tau^{m+1} e^{-s}}{\Gamma(m+1)\Gamma(n+1)}\int_0^\infty\!\!\! db\> b^m (s+b)^n e^{-b(\tau+1)}\nonumber \phantom{xxx}&&
\\
= \frac{\tau^{m+1} e^{-s}}{\Gamma(m+1)} \sum_{k=0}^n \frac{s^k}{k!\>(n-k)!} 
\frac{\Gamma(n-k+m+1)}{(\tau+1)^{n-k+m+1}}.\phantom{xxx}&&
\label{eq:DeltaPnmtaus}
\eeq
Note that here the true background mean $b$ appears only as an integration variable, 
and that 
\beq
\sum_{n=0}^\infty \Delta P(n,m,\tau,s) = 1,
\eeq 
for any $m,\tau, s$.
The limit
$\lim_{\tau \rightarrow \infty} \Delta P(n,m,\tau,s)$, with $m/\tau = \hat b$ held fixed, recovers the Poisson distribution
$P(n| s+ \hat b)$.
In the second equality of eq.~(\ref{eq:DeltaPnmtaus}), we have written a form valid for non-integer $n$ and $m$, both to define $Z^{\rm A}$ below and to account for the fact that an estimated $\hat b$ and $\Delta_{\hat b}$ may correspond to non-integer $m$. The third equality is more useful when $n$ is an integer, and also in the case $s=0$
where only the $k=0$ term survives and one can replace $n!$ by $\Gamma(n+1)$. 

The $p$-value for discovery has two equivalent forms,
\beq
&&p_{\rm disc}(n,m,\tau) \>=\> \sum_{k=n}^{\infty} \Delta P(k, m, \tau, 0)  
\nonumber 
\\ 
&&\qquad\>=\> B(1/(\tau+1), n, m+1)/B(n,m+1),\phantom{x}
\label{eq:pdiscnmtau}
\eeq
where the first form was given in \cite{Haines:1986ii,Alexandreas:1993,Linnemann:2003vw,Cousins:2008zz}
and the second (involving the ordinary and incomplete beta functions) was obtained in a frequentist approach by \cite{Gehrels:1986mj,ZhangRamsden:1990}. 
Despite appearances, these two forms are equivalent \cite{Linnemann:2003vw,Cousins:2008zz}, justifying the choice made in eq.~(\ref{eq:posteriorPb}).

For exclusion, we find\hfill
\begin{strip}
\beq
p_{\rm excl}(n,m,\tau,s) 
&=& \sum_{k=0}^n \Delta P(k,m,\tau,s) 
\,=\,
\sum_{k=0}^n \frac{\tau^{m+1}}{(\tau+1)^{k+m+1}}
\frac{\Gamma(k+m+1) \Gamma(n-k+1,s)}{k!\, \Gamma(m+1) \Gamma(n-k+1)}
\nonumber
\\
&=& \frac{\tau^{m+1}}{\Gamma(n+1) \Gamma(m+1)} \int_0^\infty\!\!\! db\> e^{-\tau b} b^m \Gamma(n+1, s+b)
\nonumber
\\
&=&
\left [\Gamma(n+1,s) - e^{-s} \int_0^\infty \!\!\! db\> e^{-b} (s+b)^n \Gamma(m+1, \tau b)/\Gamma(m+1)\right]/\Gamma(n+1)
,
\label{eq:pexclnmtau3}
\eeq
\end{strip}%
where the first form (following directly from the definition) involves a double sum, the second single-sum form is more efficient if $n$ is an integer, while the last two forms are valid for non-integer $n,m$, have differing ease of numerical evaluation depending on the inputs, and follow from each other by integration by parts.

We can now consider the expected significances in the case that $\hat b$ and $\Delta_{\hat b}$ have been fixed, corresponding either to a calculation of the background with limited accuracy, or to a measurement of $m$ for a given $\tau$. This is done by generating pseudo-experiments for $n$, distributed according to the probabilities $\Delta P(n,m,\tau,s)$ for discovery and $\Delta P(n,m,\tau,0)$ for exclusion, and then evaluating the $p$-values according to eq.~(\ref{eq:pdiscnmtau}) for discovery and eq.~(\ref{eq:pexclnmtau3}) for exclusion. As before, we consider $Z^{\rm med}$, $Z^{\rm mean}$, and $Z^{\rm A}$ obtained from the allowed pseudo-experiment data, each as functions of $s,\hat b, \Delta_{\hat b}$. Here, $Z^{\rm A}$ is obtained by replacing $n$ by its mean expected values. 
For the discovery and exclusion cases respectively, we find these are
\beq
\langle n_{\rm disc}\rangle &=& s + \widetilde b
,
\\
\langle n_{\rm excl} \rangle &=& 
\widetilde b
,
\label{eq:nmeanDeltab}
\eeq
where 
\beq
\widetilde b &=&  (m+1)/\tau = \hat b + \Delta_{\hat b}^2/\hat b .
\eeq
Then
\beq
p^{\rm Asimov}_{\rm disc}(s,\hat b, \Delta_{\hat b}) &=& p_{\rm disc}(\langle n_{\rm disc}\rangle, m, \tau) ,
\\
p^{\rm Asimov}_{\rm excl}(s,\hat b, \Delta_{\hat b}) &=& p_{\rm excl}(\langle n_{\rm excl}\rangle, m, \tau, s) 
,
\label{eq:pAsimovDeltab}
\eeq
which are converted to $Z^{\rm A}_{\rm disc}$ and $Z^{\rm A}_{\rm excl}$ as usual.

Note that the mean expected event count in the absence of signal, $\widetilde b$, is distinct from, and larger than, the measured background estimate, $\hat b = m/\tau$. The fact that 
$\widetilde b > \hat b$ can be understood heuristically as the statement that, for finite $\tau$, a given $m$ is more likely to have been a downward rather than upward fluctuation. As an extreme example, if $m=0$, this could be a downward fluctuation of a non-zero true background, but obviously it could not be an upward one. Given $(m,\tau)$, depending on the experimental situation there may be other justifiable probability density functions besides eq.~(\ref{eq:posteriorPb}), and the subsequent discussion carries through similarly for any other choice.
If we had chosen a different Bayesian distribution in eq.~(\ref{eq:posteriorPb}), then
the expression for $\widetilde b$ (in terms of $m$ and $\tau$) would change. For this reason, we prefer to
give results directly in terms of the independent variable $\hat b = m/\tau$ corresponding to the direct measurement (or calculation) of the background, rather than $\widetilde b$.    

Refs.~\cite{Cowan} and \cite{Kumar:2015tna} had earlier provided Asimov approximations to the median discovery and exclusion significances, respectively. Equations~(\ref{eq:ZAsimovCowan}) and (\ref{eq:ZAsimovKM}) above are the limits as $\Delta_b \rightarrow 0$. However, the significance estimates defined in refs.~\cite{Cowan} and \cite{Kumar:2015tna} are not directly comparable to our definitions when $\Delta_b \not=0$, since they take the (unknown) true background mean $b$ as input, rather than the point estimate $\hat b = m/\tau$ as we do here. If one ignores the distinction and considers $b = \hat b$, then $Z^{\rm A}_{\rm disc}$ and $Z^{\rm A}_{\rm excl}$ as defined in this paper give more conservative significances than those obtained from 
\cite{Cowan,Kumar:2015tna}.
\begin{figure*}
  \includegraphics[width=8.2cm]{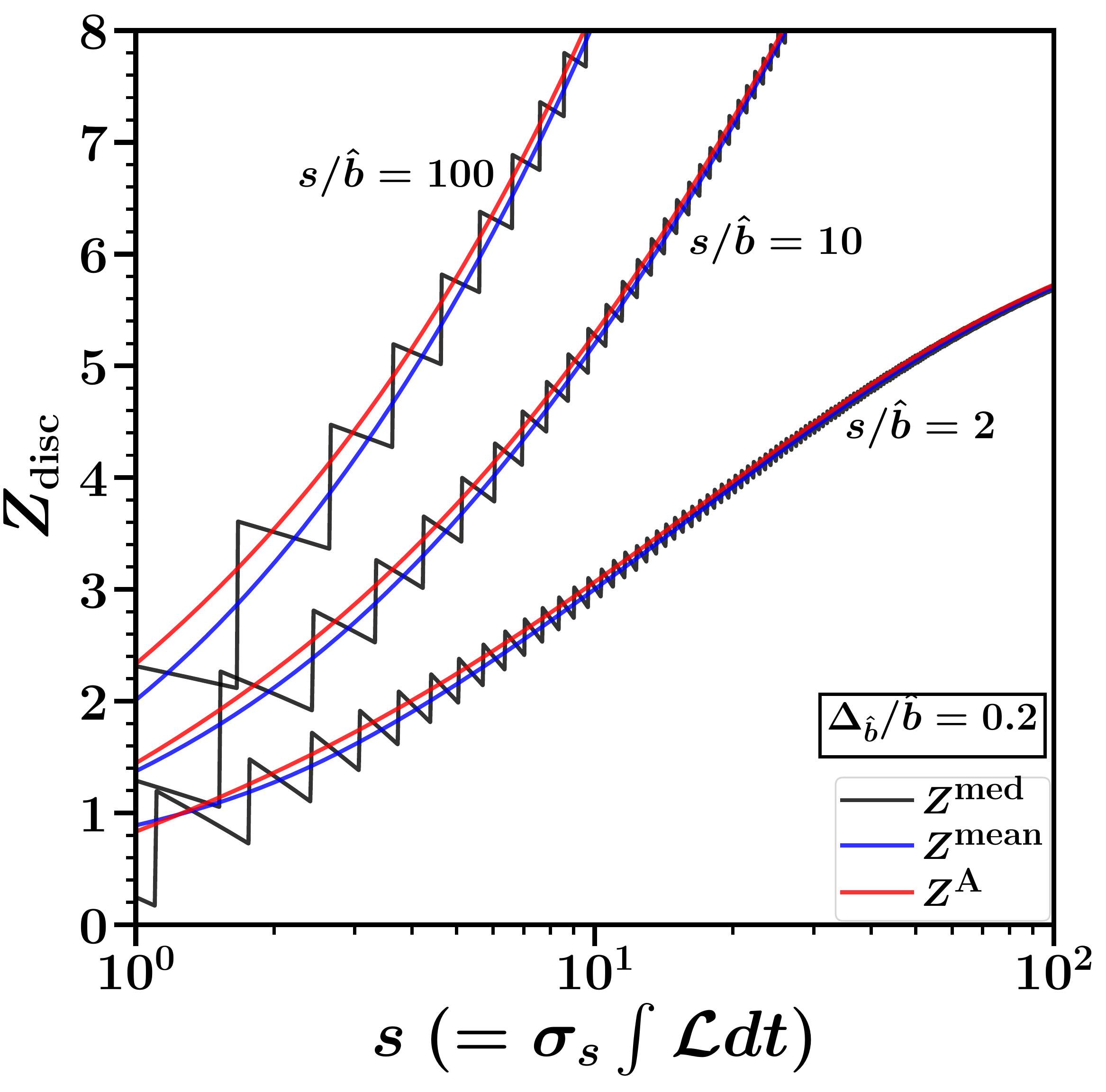}%
\hspace{0.5cm}  
  \includegraphics[width=8.2cm]{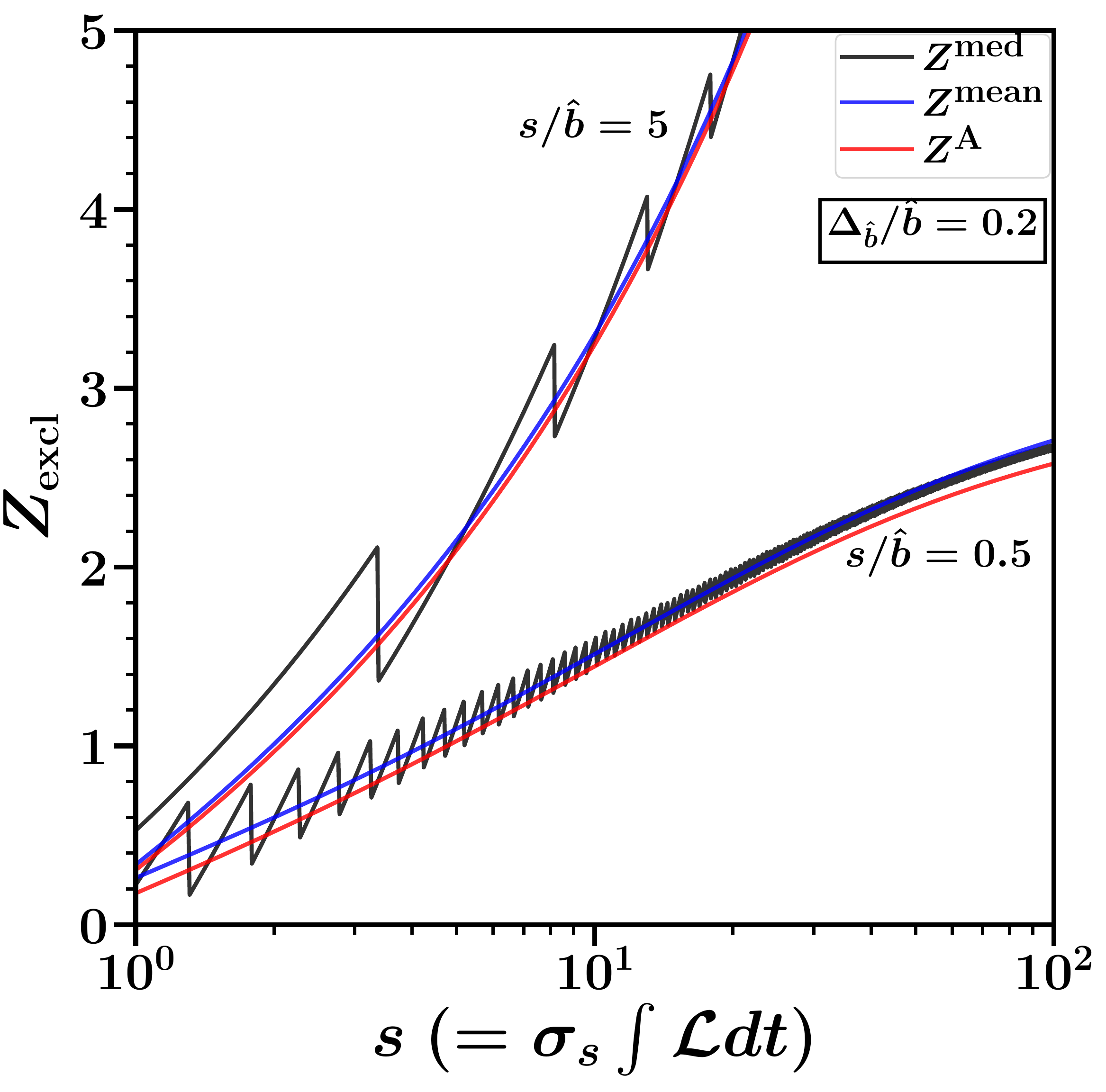}%
  \caption{The median, mean, and exact Asimov expected significances for discovery and exclusion,
  for fixed ratios $s/\hat b$ as labeled, as a function of $s$, for $\Delta_{\hat b}/\hat b = 0.2$. 
  Here $s$ and $\hat b$ are assumed to be
  proportional to their respective cross-sections multiplied by the integrated luminosity
  $\int {\cal L} dt$ of the experiment, 
  where $\sigma_s$ is the signal cross-section.
\label{fig:deltab0p2}}
\end{figure*}

\begin{figure*}
  \includegraphics[width=8.35cm]{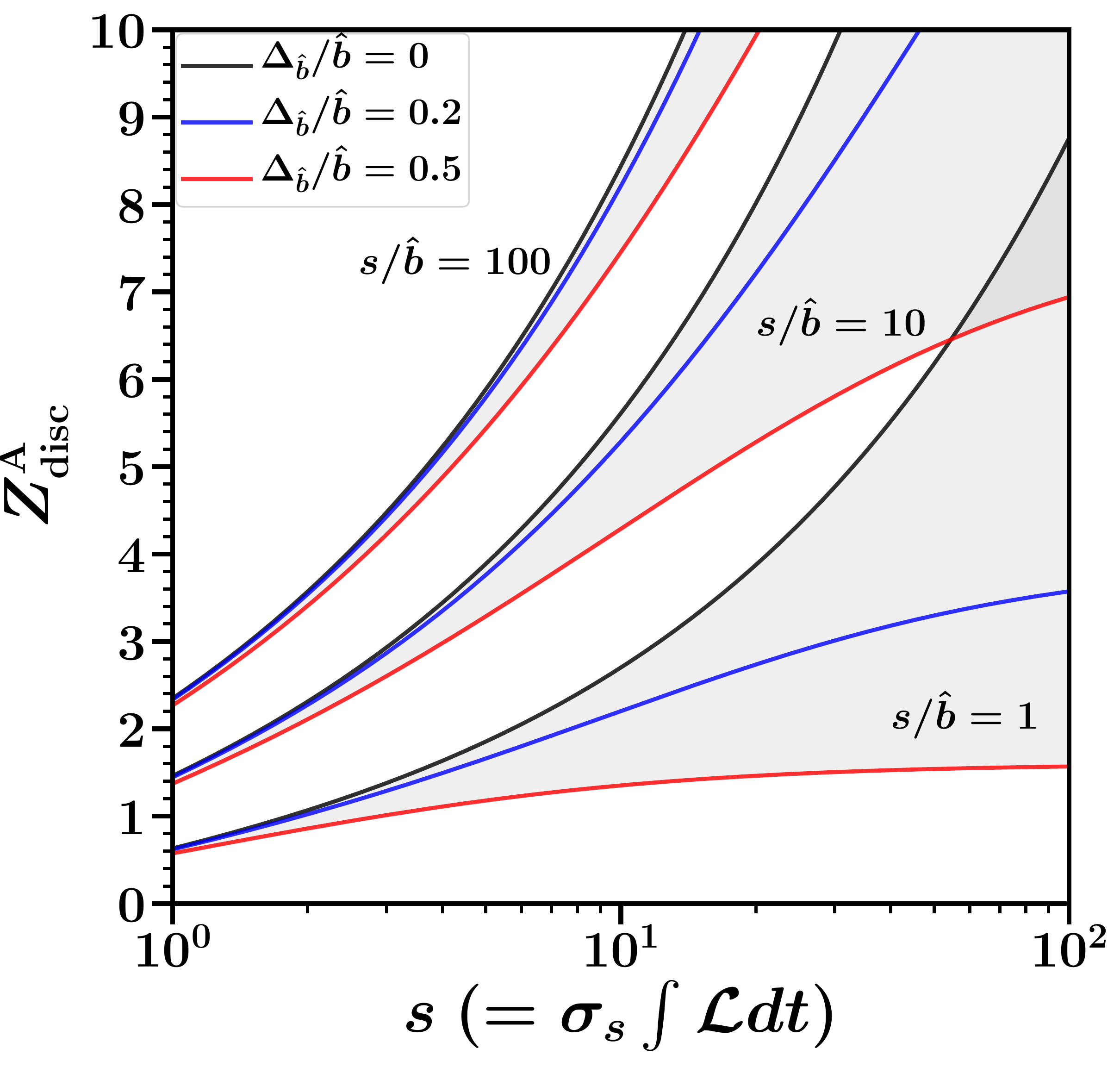}%
\hspace{0.5cm}  
  \includegraphics[width=8.1cm]{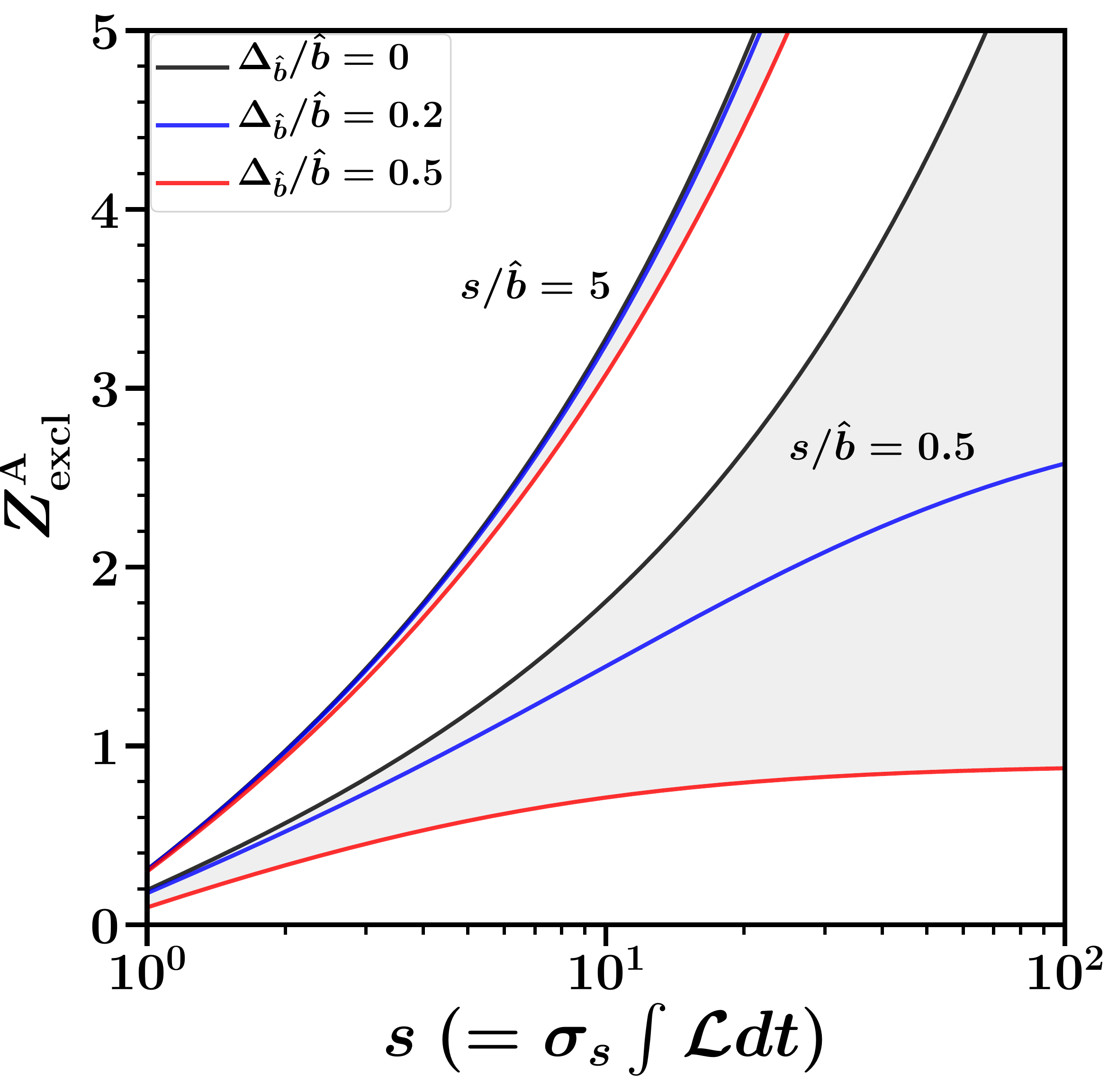}%
  \caption{The expected significance measures $Z^{\rm A}_{\rm disc}$ and $Z^{\rm A}_{\rm excl}$, 
  for fixed ratios $s/\hat b$ as labeled, as a function of $s = \sigma_s \int {\cal L} dt$, for 
  $\Delta_{\hat b}/\hat b = 0, 0.2,$ and $0.5$, as labeled. For discovery we show $s/\hat b = 1, 10, 100$,
  and for exclusion $s/\hat b = 0.5$ and $5$.
The shaded areas are the envelopes between the largest and smallest values of $\Delta_{\hat b}/{\hat b}$, for each $s/\hat b$. 
  \label{fig:ZAsimov}}
\end{figure*}

Results for $Z^{\rm med}$, $Z^{\rm mean}$, and $Z^{\rm A}$ for discovery and exclusion are shown
in Figure \ref{fig:deltab0p2} for $\Delta_{\hat b}/\hat b = 0.2$, this time for $s$ and $\hat b$ both taken proportional to
an integrated luminosity factor $\int {\cal L} dt$ which
represents the temporal progress of the experiment.
We consider fixed ratios $s/\hat b = 2, 10, 100$
for discovery and $0.5, 5$ for exclusion. Again, the sawtooth behavior of $Z^{\rm med}$ is evident, while
$Z^{\rm mean}$ and $Z^{\rm A}$ both lie within or near its envelope, and can be taken as reasonable
and monotonic measures of the expected discovery and exclusion capabilities. 
Note that $Z^{\rm A}_{\rm excl}$ is more conservative than 
$Z^{\rm med}_{\rm excl}$ or $Z^{\rm mean}_{\rm excl}$ for higher integrated luminosities, while
$Z^{\rm mean}$ is slightly more conservative for discovery. As before,
$Z^{p\>{\rm mean}}_{\rm disc} = Z^{p\>{\rm mean}}_{\rm excl}$, not shown, gives far smaller values and cannot be recommended. In Fig.~\ref{fig:ZAsimov}, we show
$Z^{\rm A}_{\rm disc}$ and $Z^{\rm A}_{\rm excl}$ for $\Delta_{\hat b}/\hat b = 0, 0.2,$ and $0.5$. 
Consistent with intuition, increasing the background uncertainty
reduces the expected significances, with a much greater impact when $s/\hat b$ is smaller.

\section{Conclusion}

In this paper, we have critically examined the use of median expected significance $Z^{\rm med}$ and 
possible alternatives. We find that
either $Z^{\rm mean}$ or $Z^{\rm A}$ as defined and evaluated above would be 
reasonable measures of the discovery and exclusion capabilities of counting experiments
with known or uncertain backgrounds. They both give results that are similar to $Z^{\rm med}$,
but are monotonic in the expected way with respect to changes in 
background and signal means and background uncertainties. They are also considerably more conservative than previous Asimov approximations, especially when the background is small. The exclusion case with low event counts, where the sawtooth behavior of $Z^{\rm med}_{\rm excl}$ is particularly prominent and problematic,
is noteworthy, as the success of the Standard Model of particle physics suggests the future importance of
limit-setting capabilities for experimental signals with small rates including rare decays, non-standard interactions, new heavy particle production, and dark matter searches.
In this paper, we have not considered the effects of uncertainty in the number of predicted signal events; this could be an interesting and important subject of future investigations.

In comparing $Z^{\rm mean}$ and $Z^{\rm A}$, we note that there is no ``correct" measure of the expected significance, since the various 
$Z$ definitions are simply different answers to different questions. 
The $Z^{\rm A}$ measure is typically slightly less conservative
in evaluating discovery, and more conservative for exclusion prospects, than $Z^{\rm mean}$. 
It may be simpler to extend $Z^{\rm A}$ to the case of experiments 
that feature more complex statistics than just integer counts of events.
Also, the $Z^{\rm A}$ measure, based on the means of the data distributions, 
is not harder to evaluate than other estimates of $Z$, provided that the probability distributions are known analytically or numerically.
In the counting experiments considered here, the evaluations of $Z^{\rm A}_{\rm disc}$
and $Z^{\rm A}_{\rm excl}$ 
require only
directly plugging into eqs.~(\ref{eq:pAsimovdiscDeltabeqzero})-(\ref{eq:pAsimovexclDeltabeqzero}) for a known background, or
eqs.~(\ref{eq:tradebhatDeltamtau}) and (\ref{eq:pdiscnmtau})-(\ref{eq:pAsimovDeltab}) for an uncertain background. For these reasons, we advocate that $Z^{\rm A}$ be the standard significance measure for projected exclusions and discovery sensitivities in counting experiments.


\section*{Appendix: supplementary material}

\begin{figure*}
 \includegraphics[width=7.9cm]{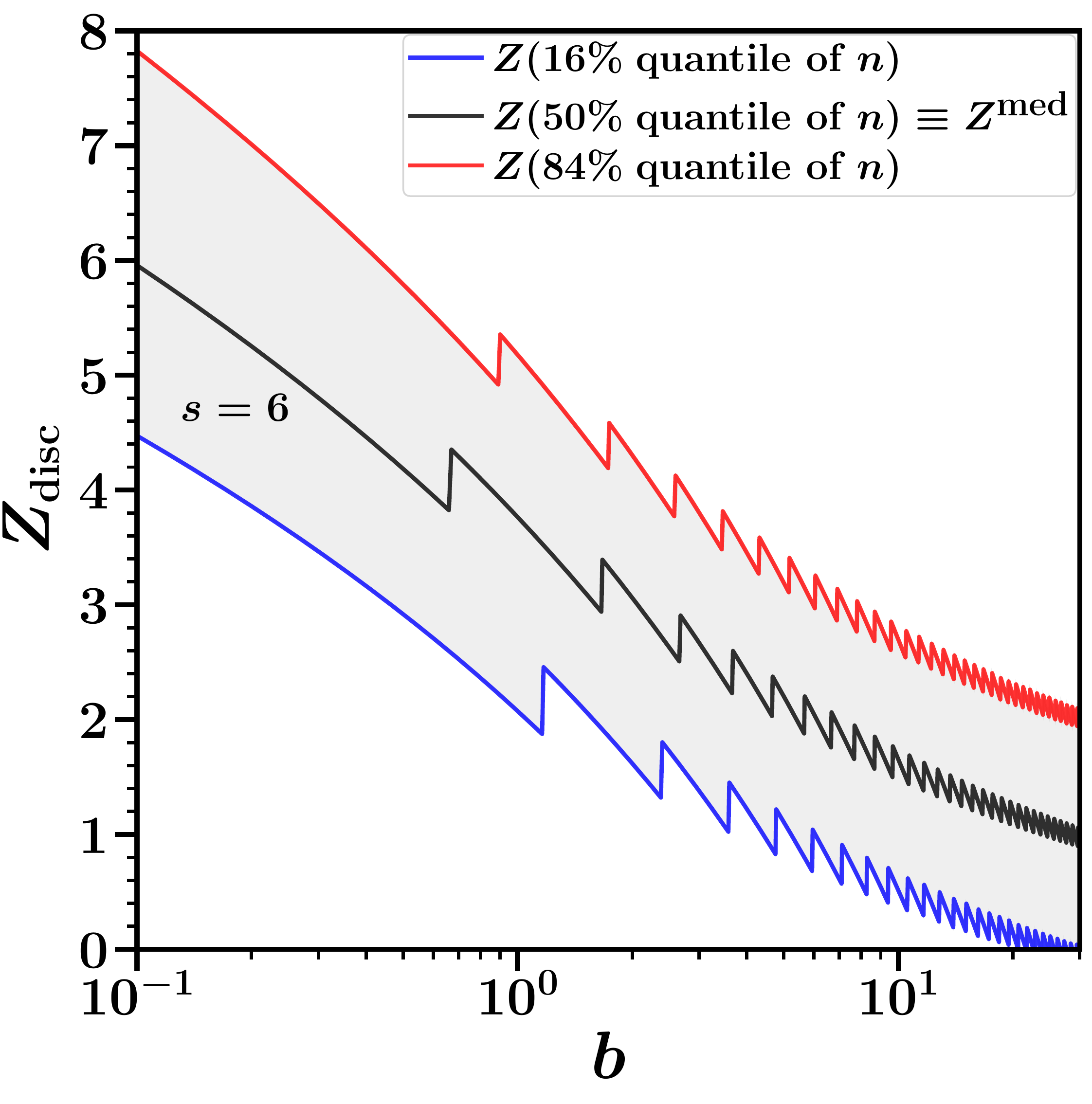}%
  \hspace{0.7cm}
 \includegraphics[width=8.15cm]{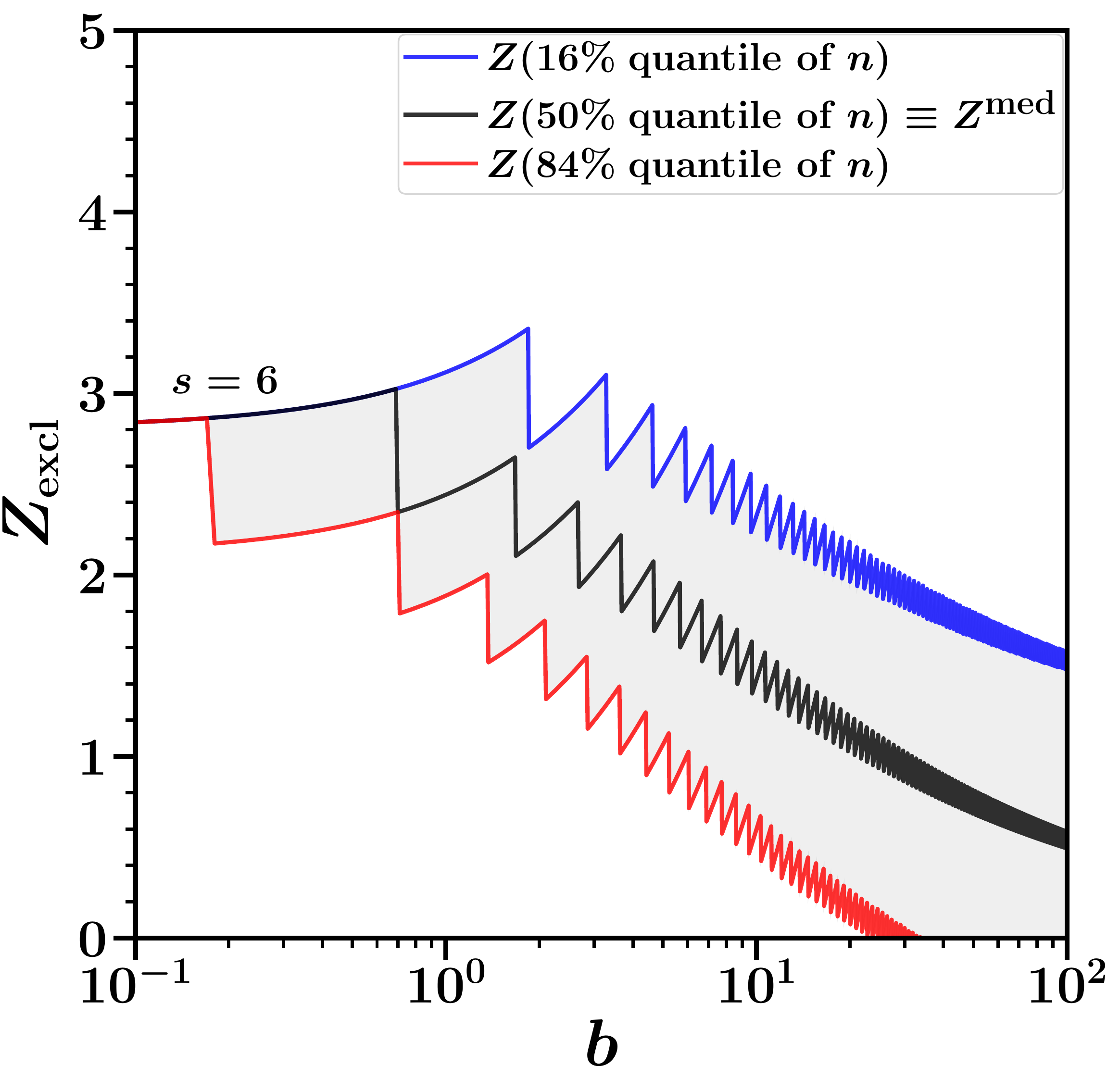}%
\caption{The 16\%, 50\%, and 84\% quantile lines for the number of pseudo-experiments $n$, for discovery (left) and exclusion (right), for signal mean $s = 6$, as functions of the background mean $b$. 
The 50\% quantile line is the median expected significance, as in Figure \ref{fig:deltabeq0}. The shaded regions are the
$1\sigma$ quantile bands.
\label{fig:quantilebands}}
\end{figure*}

To give an idea of the spread of significances for the pseudo-experiments shown in Figures
\ref{fig:deltabeq0}, we consider the $1\sigma$ bands
for each of $Z_{\rm disc}$ and $Z_{\rm excl}$, bounded by 
16\% and 84\% quantile lines for the number of pseudo-experiments $n$, 
in Figure \ref{fig:quantilebands}. Here, we took the case of signal mean $s=6$.
The 50\% quantile line is of course just the
median expected significance.

As noted above, in the case where the background estimate is determined by the method of measuring $m$ in the ``off region" and translating it to the ``on region" through $\tau$, it is possible to consider different
Bayesian priors for the true background mean $b$, rather than the flat prior chosen in the main text. For a simple two-parameter class of examples, consider
\beq
{\rm Prior}(b) \,&\propto&\, b^q e^{-\theta b},
\eeq
where $q = \theta = 0$ recovers the choice made in the main text. 
Then one finds a normalized Bayesian posterior distribution
for the background, in place of eq.~(\ref{eq:posteriorPb}):
\beq
P(b| m,\tau) &=& (\tau + \theta)^{m+q+1} b^{m+q} e^{-b(\tau + \theta)}/\Gamma(m+q+1) .
\nonumber \\ 
&&
\eeq
The calculations of $\Delta P$, $p_{\rm disc}$, and $p_{\rm excl}$ would then go through as before with
the replacements $\tau \rightarrow \tau+\theta$ and $m \rightarrow m+q$, with the results still 
expressible in terms of the independent variables
$\hat b$ and $\Delta_{\hat b}$ as defined by eq.~(\ref{eq:tradebhatDeltamtau}).
In particular, one would have 
\beq
\widetilde b &=& (m+q+1)/(\tau + \theta)\\ 
&=& \hat b [1 + (q+1) \Delta_{\hat b}^2/\hat b^2]/[1 + \theta \Delta_{\hat b}^2/\hat b]
\eeq 
in that case. However, in the absence of a compelling reason to the contrary,
we consider the simple flat prior $q = \theta = 0$ to be preferred, as it successfully reproduces the frequentist result eq.~(\ref{eq:pdiscnmtau}) for $p_{\rm disc}$, as shown in \cite{Linnemann:2003vw,Cousins:2008zz}. In any case, the $Z^{\rm mean}$ and $Z^A$ measures can be defined as above with any suitable choice of prior as dictated by realistic considerations.

\begin{figure*}[h]
  \includegraphics[width=8.4cm]{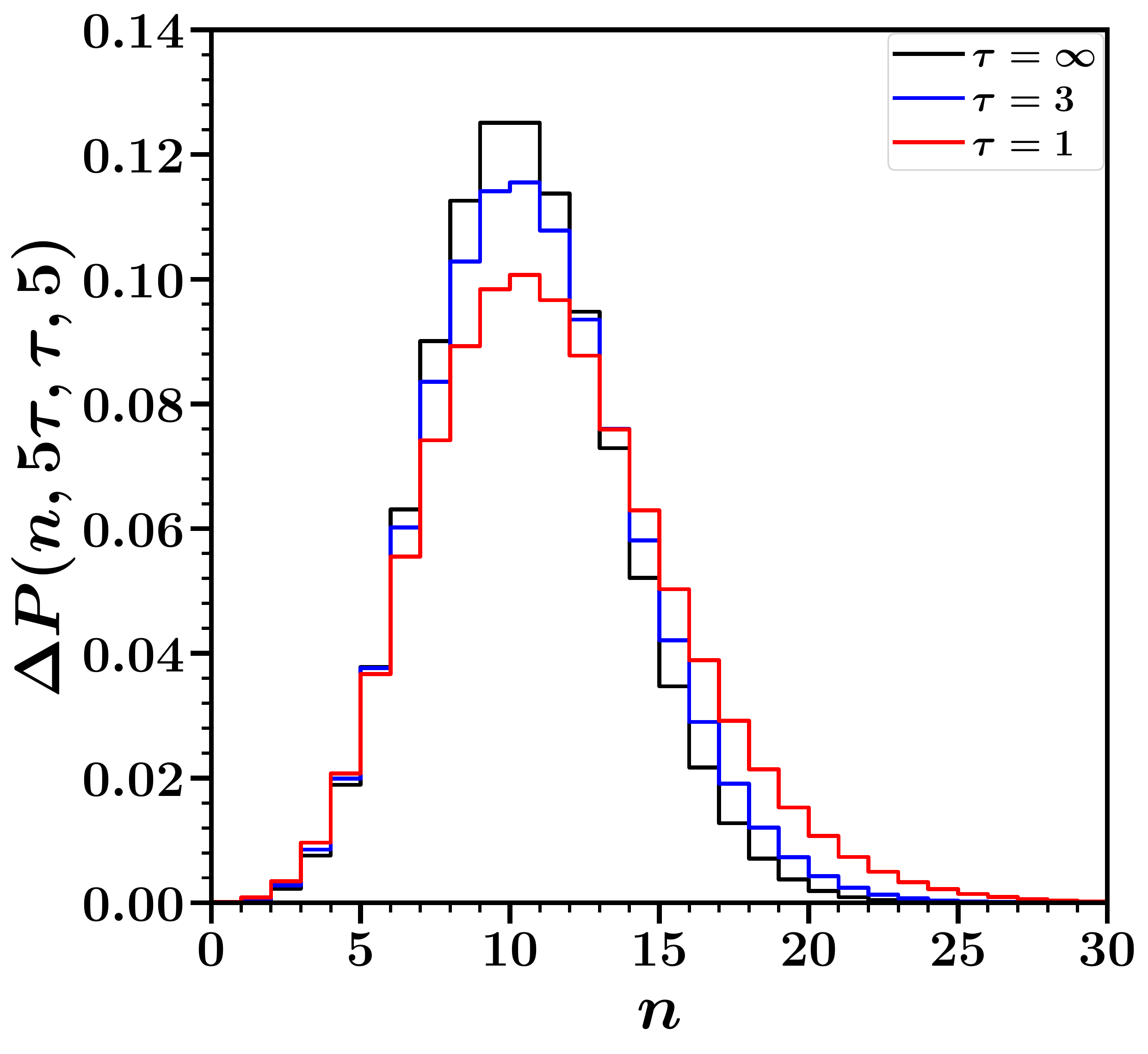}%
 \hspace{0.5cm}  
  \includegraphics[width=8.4cm]{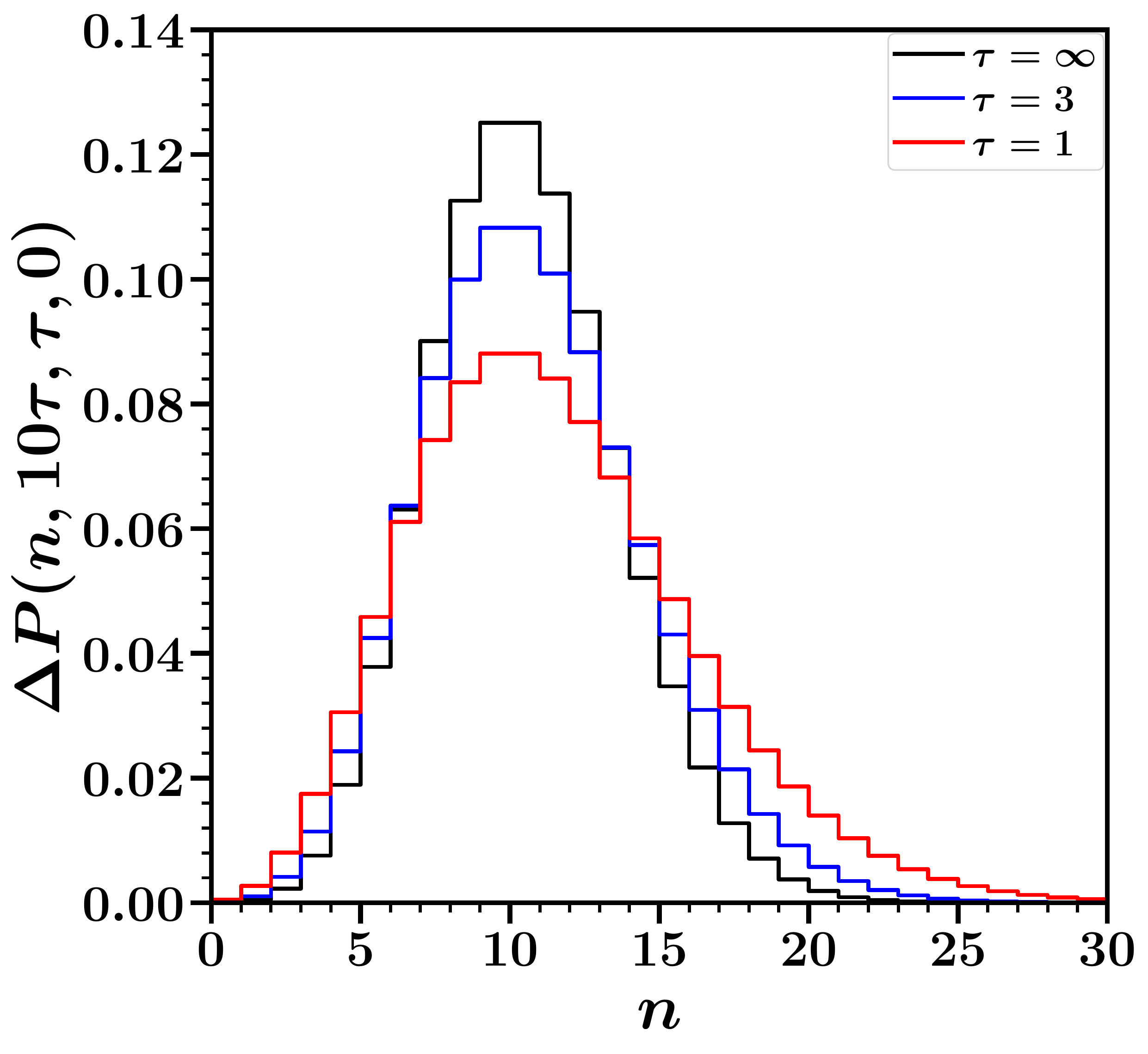}%
  \caption{
  The distributions $\Delta P(n,m,\tau,s)$, for $s=5$, $\hat b =m/\tau= 5$
  (left panel) and $s=0$, $\hat b =m/\tau= 10$ (right panel), for $\tau = 1, 3$, and $\infty$.
  In each case, 
  the result for $\tau=\infty$ is
  the Poisson distribution $P(n|s+\hat b) = P(n|10)$.
  \label{fig:DeltaP}}
\end{figure*}

We now show some further results supplementary to our main discussion.
In Fig.~\ref{fig:DeltaP}, we first show the probabilities $\Delta P (n, m, \tau, s)$
for discovery (left panel) and $\Delta P (n, m, \tau, 0)$ for exclusion (right panel),
for a fixed $\hat b = m/\tau$, as a function of event count $n$ in the signal (on) region,
for various values of $\tau$. The lines for $\tau = \infty$ in both panels correspond to the Poisson
distribution $P(n|\mu)$ with $\mu=s+\hat b$ for the discovery case, and $\mu=\hat b$ for the exclusion case. For a fixed $\hat b$, as $\tau$ gets larger, the
$\Delta P$ distribution approaches the Poisson distribution, as expected.

\begin{figure*}[h]
  \includegraphics[width=8.4cm]{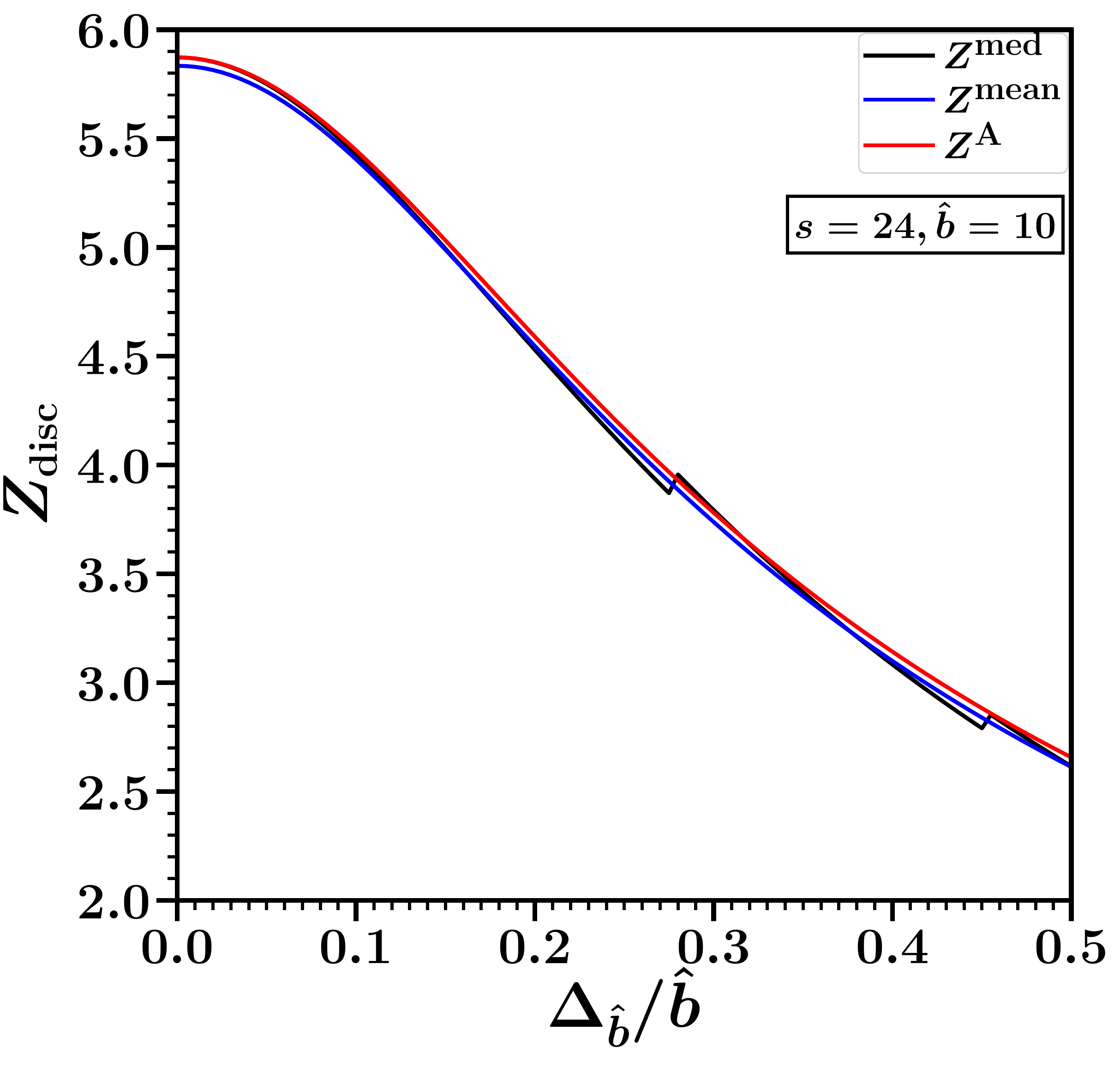}%
 \hspace{0.5cm}  
  \includegraphics[width=8.4cm]{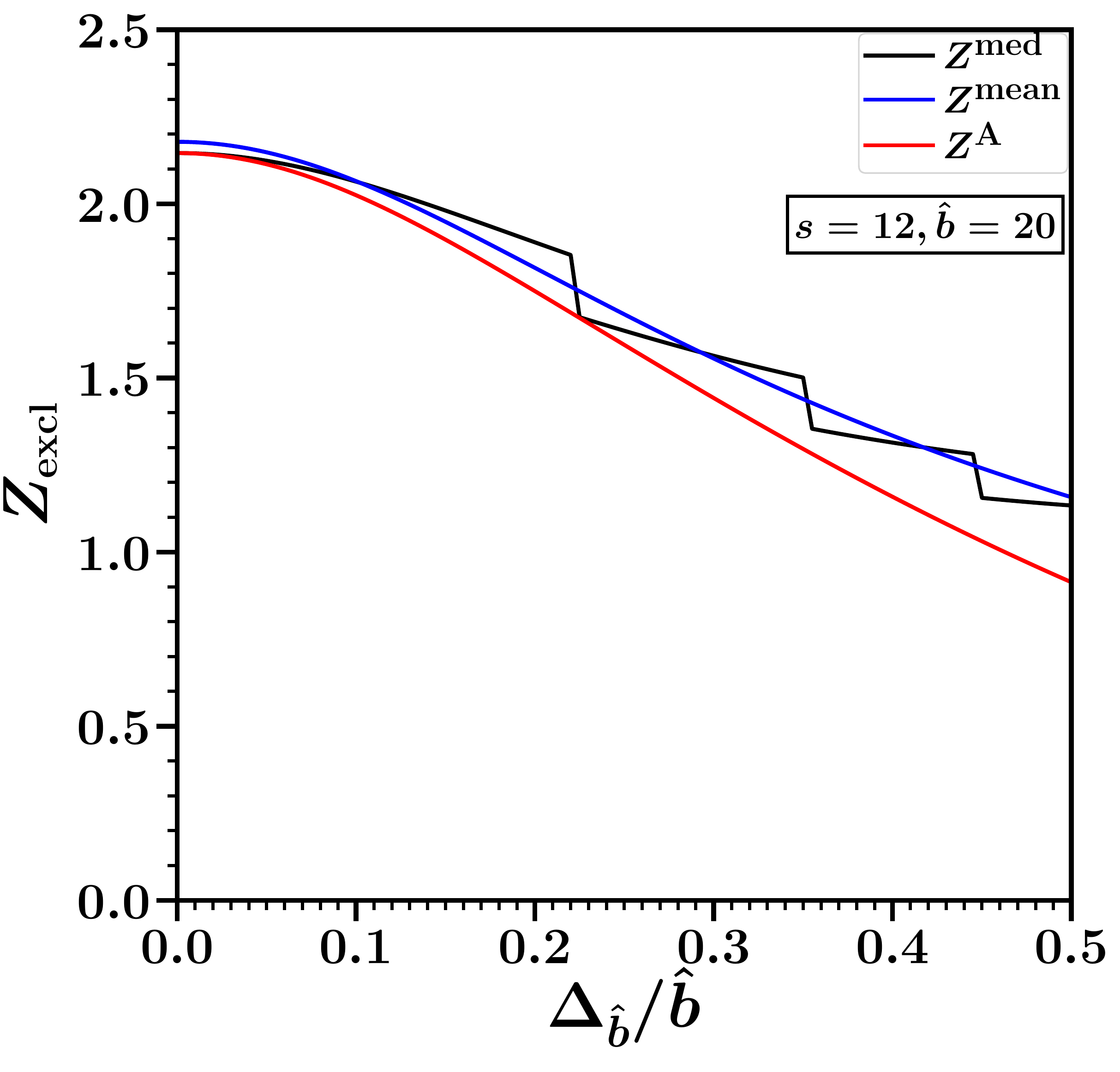}%
  \caption{
  The median, mean, and exact Asimov expected significances for discovery with 
  $s=24, \hat b=10$ (left) and exclusion with $s=12, \hat b = 20$ (right),
  as a function of $\Delta_{\hat b}/\hat b$.
  \label{fig:Zdeltabob}}
\end{figure*}

Intuitively, we also expect the discovery and exclusion significance measures
to dramatically decrease when the background uncertainty gets larger.
From Fig.~\ref{fig:Zdeltabob}, we see that the median expected
significance, once again, suffers from the sawtooth behavior.
However, the expected significances $Z^{\rm mean}$ and $Z^{\rm A}$
behave as we expect, and, as argued above, can be taken as
reasonable measures of the expected discovery and exclusion significances.
Also, it is evident from the figure that the
$(\Delta_{\hat b}, \hat b) \rightarrow (0, b)$ limit works out smoothly.

\begin{figure*}
  \includegraphics[width=8.4cm]{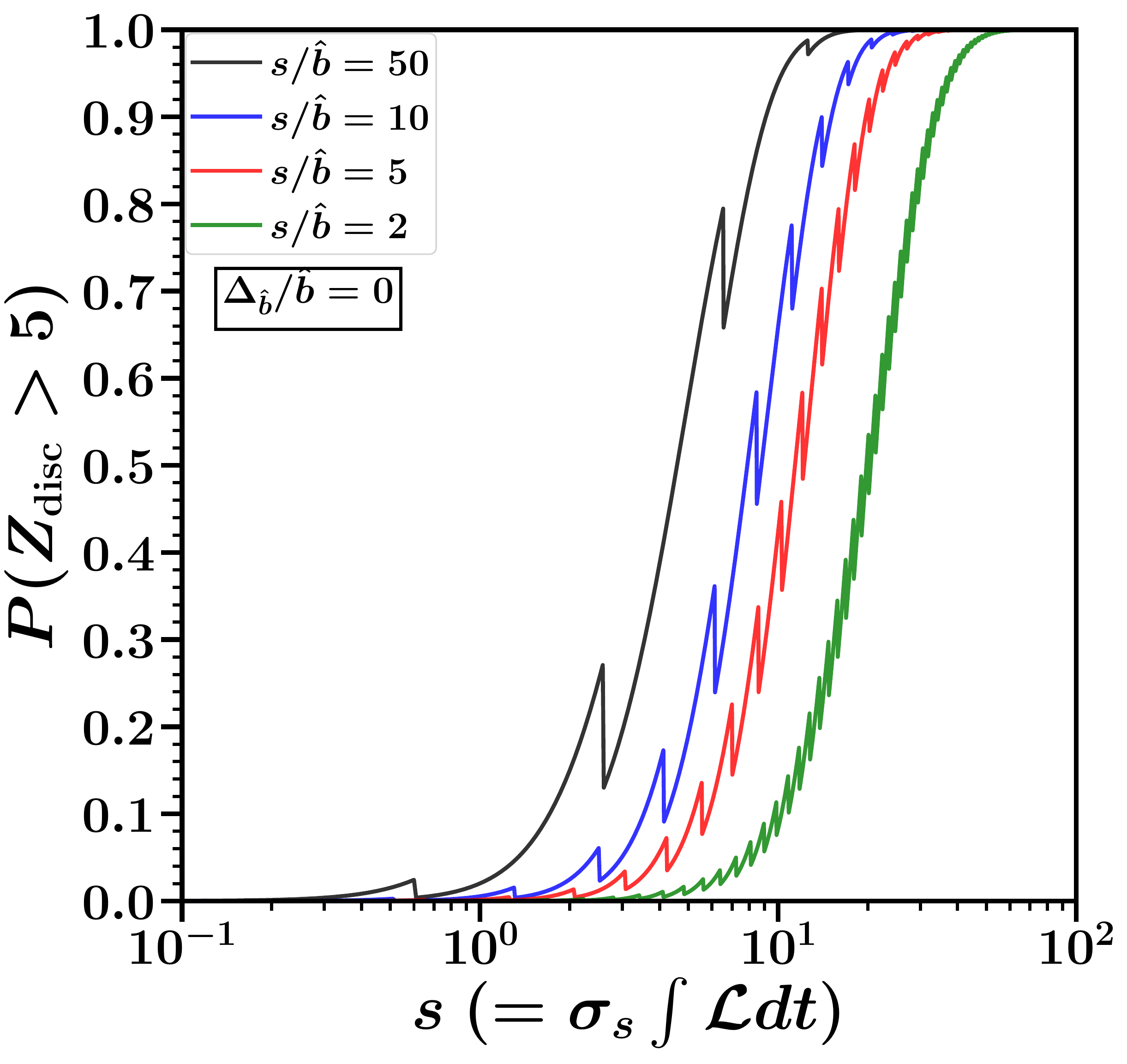}%
 \hspace{0.5cm}  
  \includegraphics[width=8.4cm]{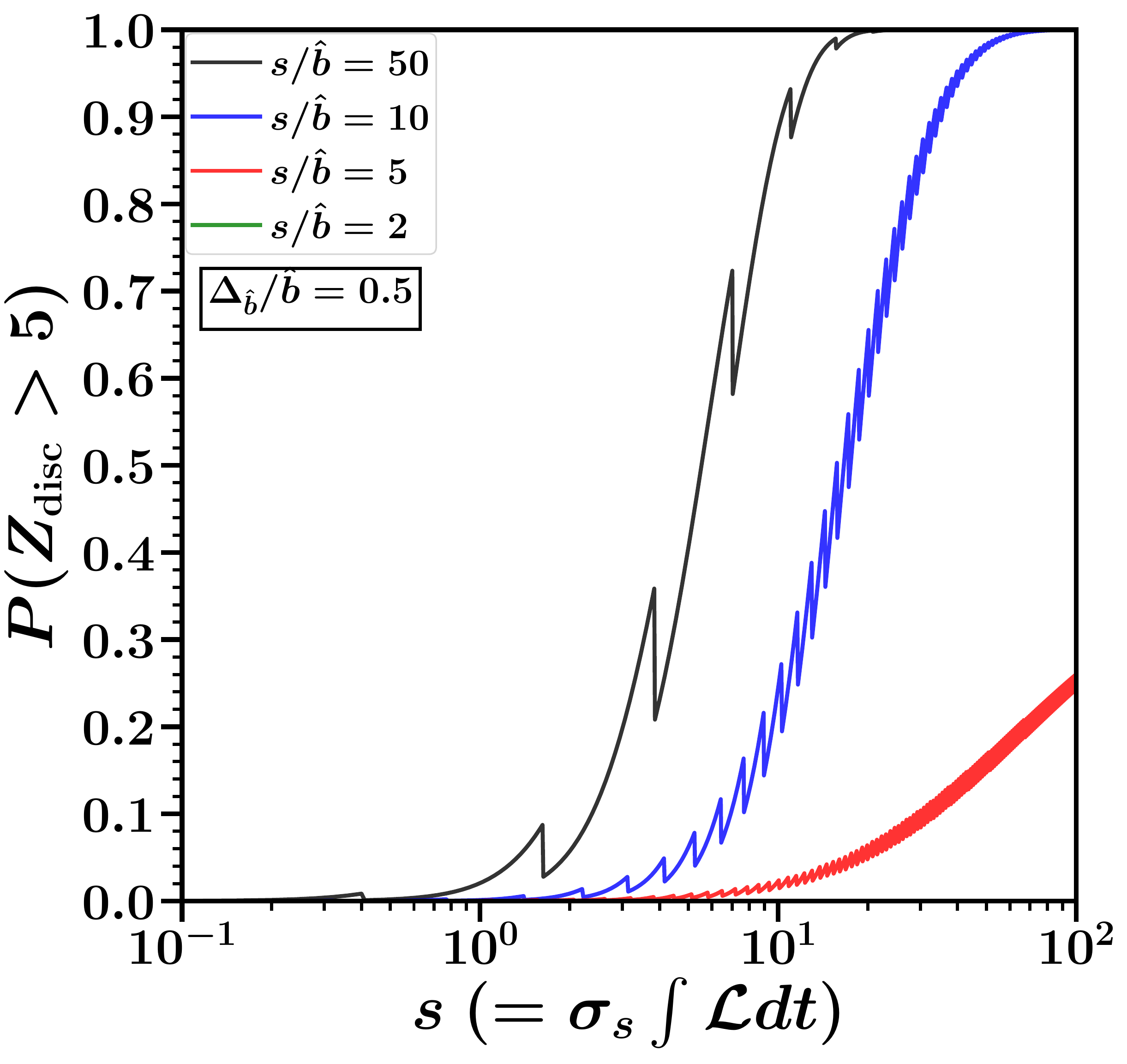}%
  \caption{
  The probability of obtaining a significance $Z_{\rm disc} > 5$, corresponding
  to greater than $5 \sigma$ discovery, in a large number of pseudo-experiments generated
  for the discovery case, for fixed ratios $s/\hat b = 2, 5, 10,$ and $50$,
  as a function of $s = \sigma_s \int {\cal L} dt$, for $\Delta_{\hat b}/\hat b = 0$ (left)
  and $\Delta_{\hat b}/\hat b = 0.5$ (right).
  \label{fig:pZdisc}}
\end{figure*}

\begin{figure*}[!h]
  \includegraphics[width=8.4cm]{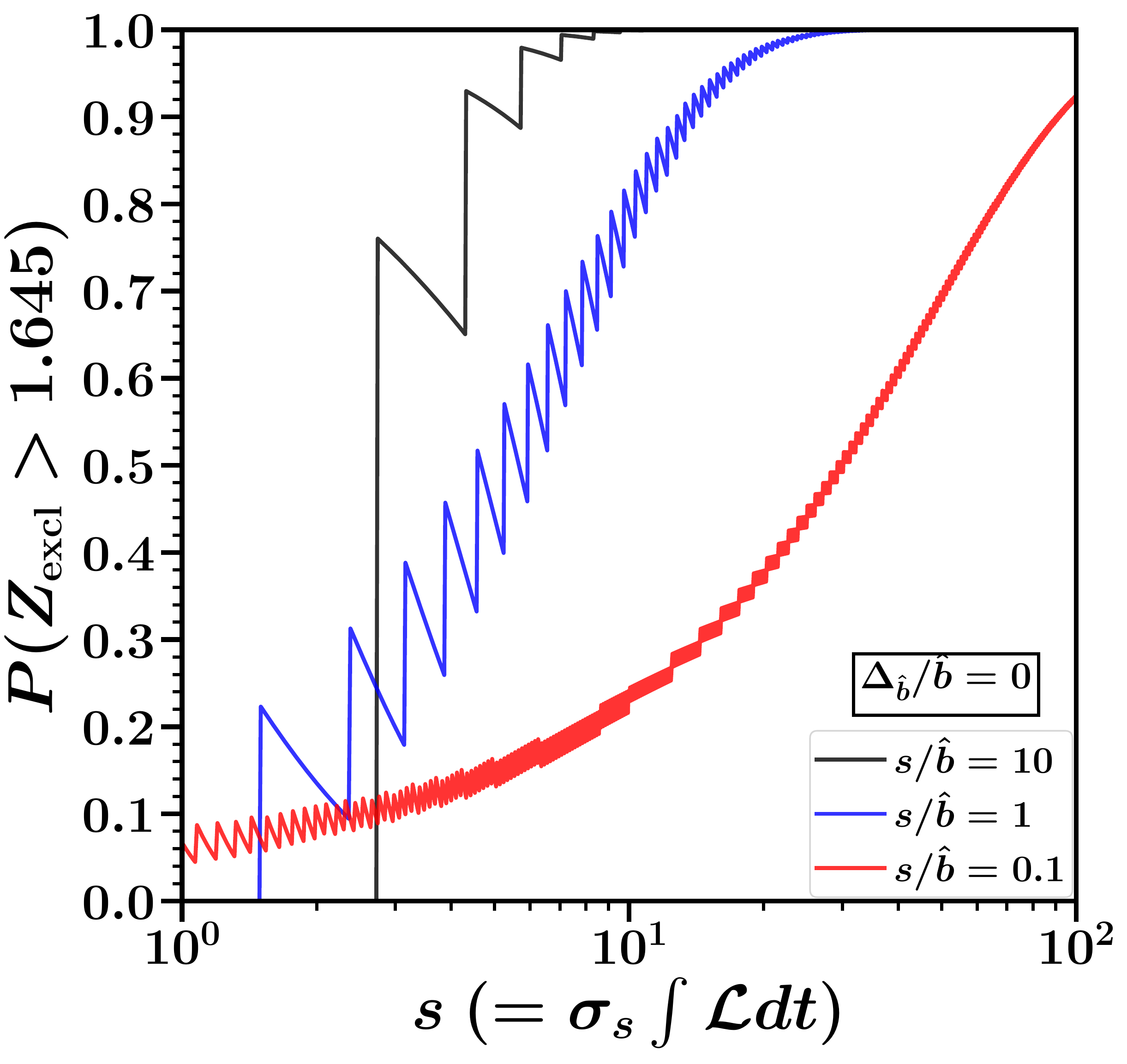}%
 \hspace{0.5cm}  
  \includegraphics[width=8.4cm]{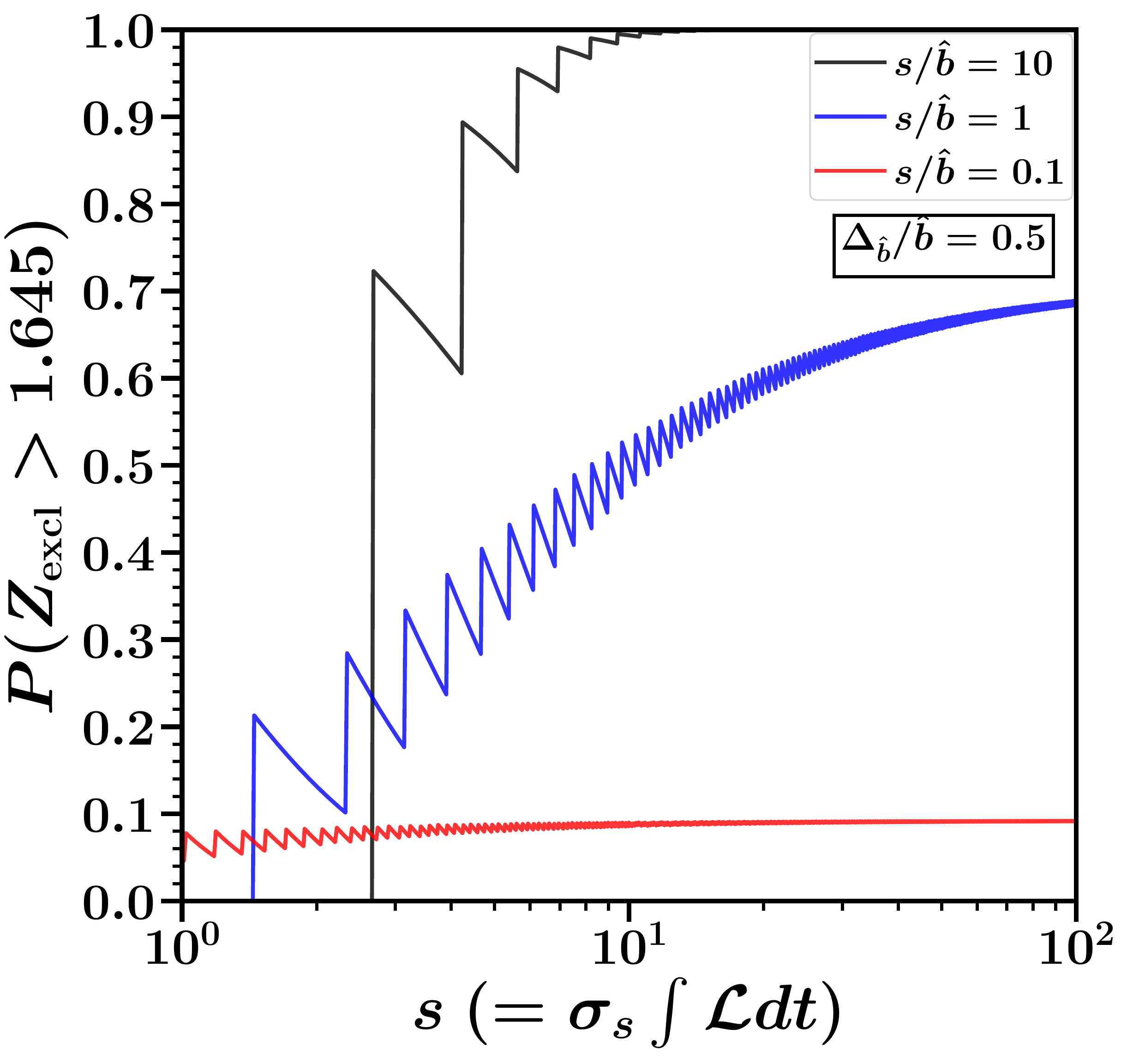}%
  \caption{
  The probability of obtaining a significance $Z_{\rm excl} > 1.645$, corresponding
  to greater than $95\%$ CL exclusion, in a large number of pseudo-experiments generated
  for the exclusion case, for fixed ratios $s/\hat b = 0.1, 1,$ and $10$,
  as a function of $s = \sigma_s \int {\cal L} dt$, for $\Delta_{\hat b}/\hat b = 0$ (left)
  and $\Delta_{\hat b}/\hat b = 0.5$ (right).
  \label{fig:pZexcl}}
\end{figure*}

With the motivation of considering other statistical measures,
we now examine some alternatives to the median, mean, or Asimov expected $Z$.
For a large number of pseudo-experiments simulated for the discovery case, we can also count the number
of these experiments, where we have greater than $5 \sigma$ discovery, and thus obtain
a probability $P(Z_{\rm disc} > 5)$. In Fig.~\ref{fig:pZdisc}, we compare $P(Z_{\rm disc} > 5)$
for $\Delta_{\hat b}/\hat b = 0$ (left panel), and 0.5 (right panel). As we expect, $P(Z_{\rm disc} > 5)$
decreases, more drastically for smaller $s/\hat b$, as the background uncertainty increases.
However, this measure also shows a sawtooth behavior, rather than increasing monotonically with
$s = \sigma_s \int {\cal L} dt$.
Similarly, Fig.~\ref{fig:pZexcl} shows the probability of obtaining greater than $95\%$ CL exclusion
in a large number of pseudo-experiments simulated for the exclusion case $P(Z_{\rm excl} > 1.645)$
for $\Delta_{\hat b}/\hat b = 0$ (left panel), and 0.5 (right panel). Once again, increasing the background
uncertainty reduces $P(Z_{\rm excl} > 1.645)$, more drastically for smaller $s/\hat b$. And, as was
the case with $P(Z_{\rm disc} > 5)$, this measure also shows a sawtooth behavior with
respect to changes in $s = \sigma_s \int {\cal L} dt$.

\begin{figure*}[!h]
  \includegraphics[width=8.4cm]{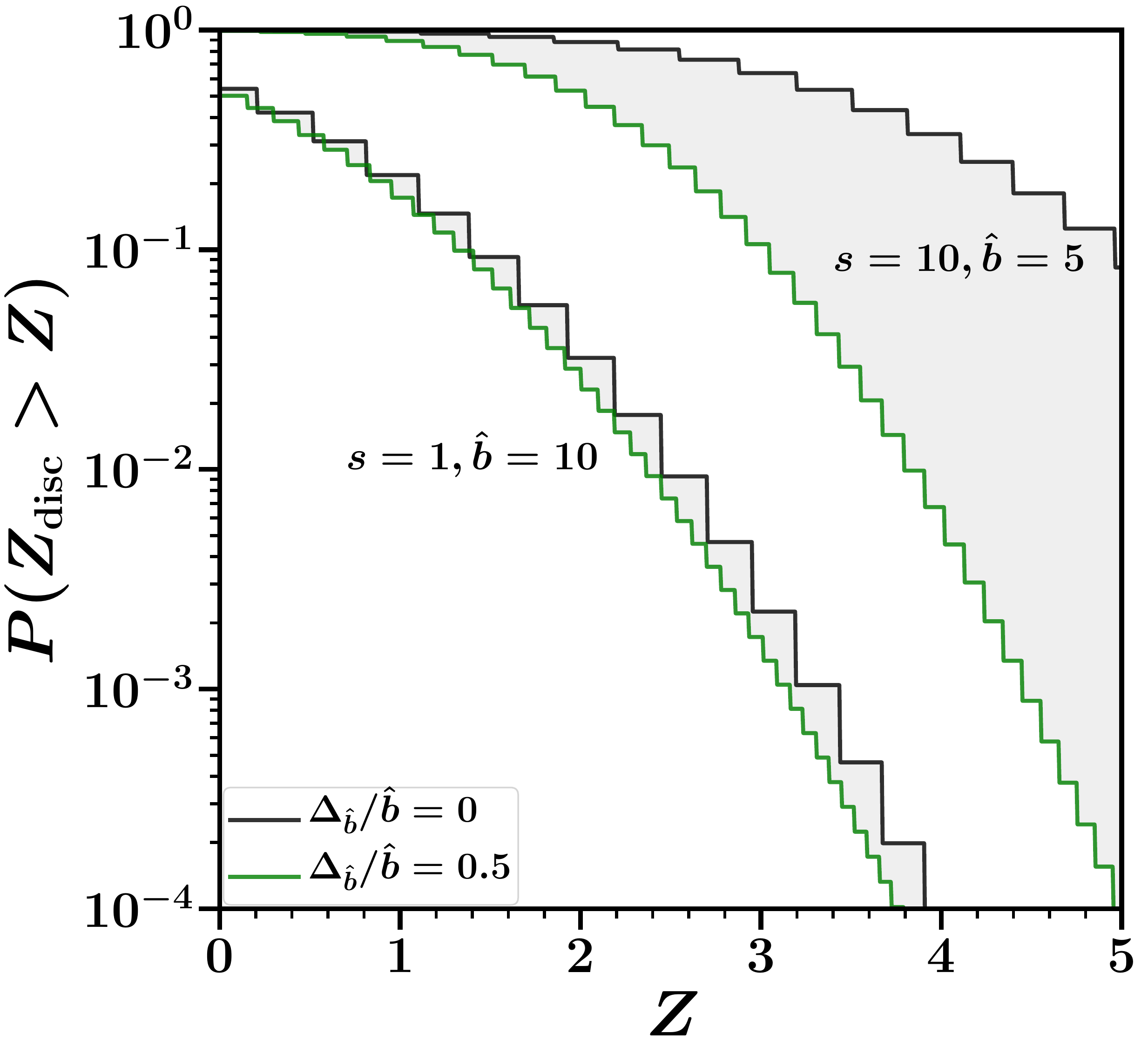}%
 \hspace{0.5cm}  
  \includegraphics[width=8.4cm]{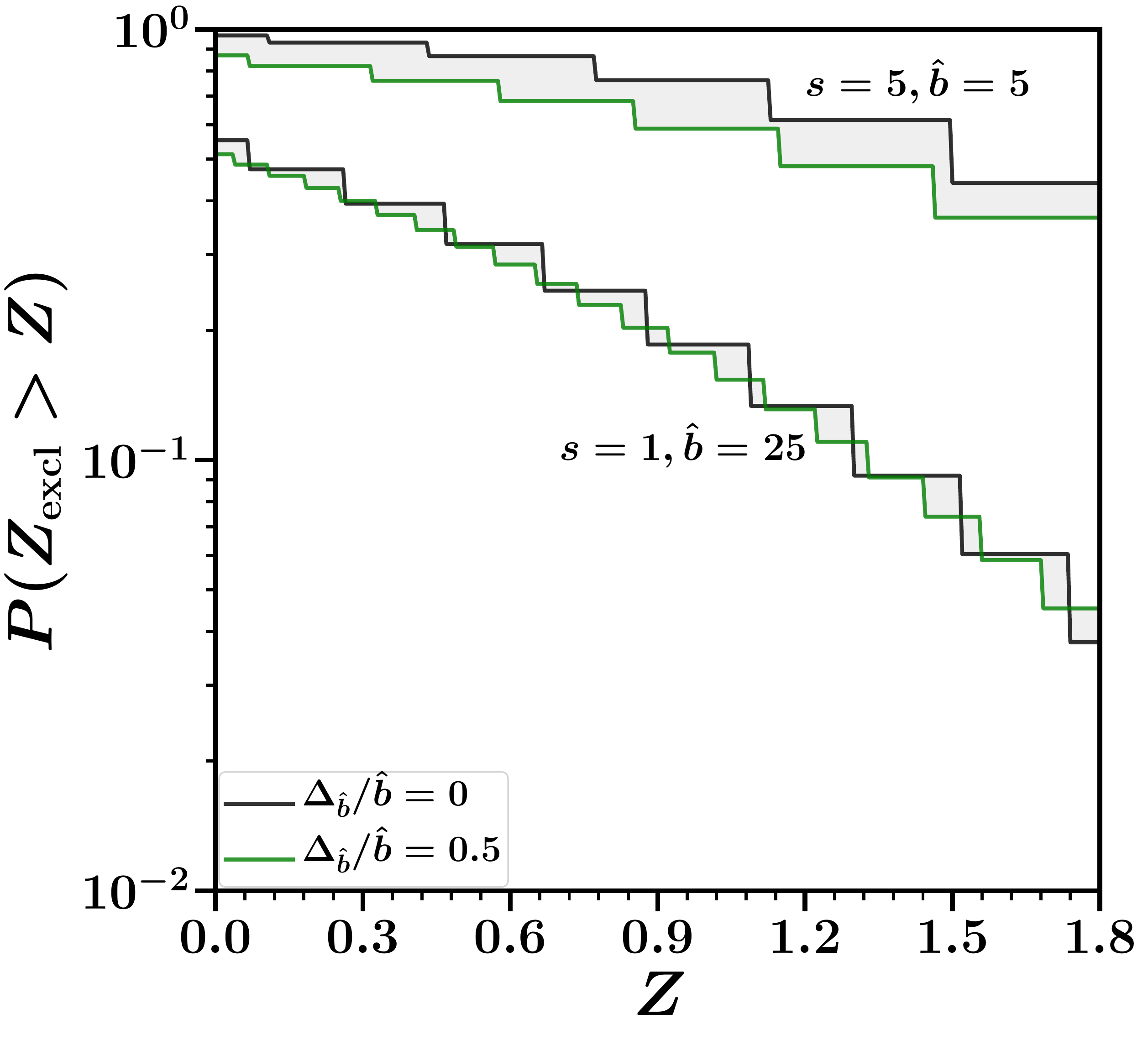}%
  \caption{
  The probability of obtaining a discovery significance $Z_{\rm disc} > Z$ (left)
  and an exclusion significance $Z_{\rm excl} > Z$ (right) in a large number
  of pseudo-experiments, for various $(s, \hat b)$ as labeled,
  for $\Delta_{\hat b}/\hat b = 0,$ and $0.5$, as a function of $Z$.
  \label{fig:pZvsZ}}
\end{figure*}

Finally, Fig.~\ref{fig:pZvsZ} shows the probability of obtaining a significance greater than a certain $Z$
in a large number of pseudo-experiments simulated for both discovery (left panel) and
exclusion (right panel) cases, for fixed $(s, \hat b)$ and $\Delta_{\hat b}/\hat b = 0, 0.5$, as a function of $Z$.
As expected, both $P(Z_{\rm disc} > Z)$ and $P(Z_{\rm excl} > Z)$ decrease with increasing $Z$,
and with increasing background uncertainty.
However, for smaller $s/\hat b$, background uncertainty does not have much impact on the results. 

A Python implementation of various significance measures for projected exclusions and discovery sensitivities
in counting experiments examined in this letter, including the advocated $Z^{\rm A}$, is made available in a code repository
{\sc Zstats} at:\\
\url{https://github.com/prudhvibhattiprolu/Zstats} .\\
To illustrate the usage of the code, the repository also has short programs that produce the data in each of the figures in this paper.
More information about all functions in this package can also be accessed using the Python help function.

\medskip
\begin{acknowledgements}
This work is supported in part by the National Science Foundation under grant numbers 1719273 and 2013340.
This work was supported in part by the Department of Energy (DE-SC0007859).
\end{acknowledgements}


\end{document}